\documentclass[journal,10pt]{IEEEtran}
\usepackage[utf8]{inputenc}
\usepackage{graphicx} 
\usepackage{booktabs} 
\usepackage{amsmath}
\usepackage{amsthm}
\usepackage{bm}
\usepackage{gensymb}
\usepackage{physics}
\usepackage{mathtools, nccmath}
\usepackage{algorithm}
\usepackage{algpseudocode}
\usepackage{xcolor}
\usepackage{setspace}
\usepackage{amsfonts}
\usepackage{amssymb}
\usepackage{comment}
\usepackage{romannum}
\usepackage[center]{caption}
\usepackage[justification=justified,skip=0pt]{caption}
\usepackage{bbm}
\usepackage{cite}
\usepackage{lipsum}
\usepackage[english]{babel}
\usepackage{amsthm}
\usepackage{authblk}
\usepackage{adjustbox}
\usepackage{color}
\usepackage{comment}

\allowdisplaybreaks
\usepackage[left=0.625in, right=0.625in, top=0.75in, bottom=1in]{geometry}
\IEEEoverridecommandlockouts

\title{Transceiver Design for  Cell-Free \\ Unsourced Random Access}

\author[*]{Mert Ozates\thanks{Part of this work is presented at the 2025 IEEE 101st Vehicular Technology Conference (VTC2025-Spring) \cite{vtc}.}\thanks{Mert Ozates and Eduard Jorswieck acknowledge the financial support by European Commission’s Horizon Europe, Smart Networks and Services Joint Undertaking, research and innovation program under grant agreement number 101139282, 6G-SENSES project, and by the German Federal Ministry of Research, Technology and Space (BMFTR) within the project xG-RIC under grant reference number 16KIS2433 as part of the research program Communication Systems ''Souverän. Digital. Vernetzt.''. 
}}
\author[**]{Mohammad Kazemi}
\author[***]{Eduard Jorswieck}
\author[**]{Deniz Gündüz}
\affil[*]{IHP - Leibniz Institute for High Performance Microelectronics, 15236 Frankfurt (Oder), Germany \protect\\ Email: oezates@ihp-microelectronics.com}
\affil[**]{Department of Electrical and Electronic Engineering, Imperial College London  \protect\\ Email:\{mohammad.kazemi, d.gunduz\}@imperial.ac.uk}
\affil[***]{Institute for Communications Technology, Technische Universität Braunschweig, Germany \protect\\Email:  e.jorswieck@tu-braunschweig.de}
\date{}

\begin{document}

\pagenumbering{arabic}
\maketitle

\begin{abstract}
    We propose a low-complexity, energy-efficient solution for cell-free unsourced random access (URA), in which multiple distributed access points are connected to a central processing unit via a fronthaul. Assuming that each user transmits a pilot sequence followed by a polar codeword whose symbols are placed to the data part of the frame according to an on-off pattern, we utilize iterative decoding, including orthogonal matching pursuit (OMP)-based pilot detection and channel estimation, symbol estimation using a linear minimum mean square error (MMSE) solution, symbol combining, single-user decoding, and successive interference cancellation (SIC). We also present finite blocklength (FBL) performance and detailed complexity analyses. Numerical results demonstrate that the FBL analysis properly characterizes system performance and that the proposed cell-free URA scheme offers superior performance compared to existing low-complexity schemes. Namely, it offers a superior performance of up to 4.5 dB and can accommodate up to 1800 active users.
\end{abstract}



\section{Introduction}

6G and beyond communication systems are expected to support massive connectivity with high reliability and low latency, making massive machine-type communication (mMTC) applications (e.g., Internet of Things (IoT)) a central use case. In these applications, a massive number of devices with short payloads send their messages to a base station (BS) without any coordination in a sporadic fashion, i.e., only a small fraction of the devices are active at any given time. In \cite{polyanskiy}, this problem is treated as a coding problem in the physical layer called unsourced random access (URA) \cite{survey}, in which each device uses the same codebook. This removes the user identity, making system operation independent of the total number of devices and reducing the receiver's task to extracting a list of transmitted messages up to a permutation. Accordingly, per-user probability of error (PUPE) is adopted as the main performance metric.



In \cite{polyanskiy}, Polyanskiy also presents an achievable scheme and characterizes its energy efficiency over a real-valued Gaussian multiple-access channel (MAC). Several low-complexity energy-efficient URA schemes have since been developed based on $T$-fold slotted ALOHA \cite{ordentlich,vem,marshakov}, coded compressed sensing (CCS) \cite{comp2,comp4,nassaji}, random spreading \cite{pradhan,pradhan2,schiavone}, on-off division multiple access (ODMA) \cite{ozates,yan}, and sparse linear precoding, providing a unification of random spreading and ODMA\cite{dang}. While a synchronized setup is studied in \cite{ordentlich}-\cite{dang}, asynchronous URA is studied in \cite{wu} and \cite{wu2}, where the delay is bounded by a fraction of the transmission frame length. Instead, a fully asynchronous setup is considered in \cite{karami} and \cite{ozates2}, where the users can initiate their transmissions at any time instance without any restriction. URA over a fading MAC with a single antenna receiver is studied in \cite{ozates},\cite{andreev,nassaji2,zhang,kowshik,ozates3}. In \cite{andreev}, treating interference as noise (TIN) coupled with successive interference cancellation (SIC) is employed, while the diversity of the repeated channel-coded sequences is utilized in \cite{nassaji2}. While \cite{ozates} employs ODMA with the aid of pilot sequences, a pilot-free ODMA scheme with high energy efficiency was recently proposed in \cite{zhang}. Alternatively, orthogonal frequency division multiplexing (OFDM) is employed in the asynchronous setup in \cite{kowshik}, converting the time synchronization problem into phase-shift estimation, and in \cite{ozates3} to mitigate the detrimental effects of frequency-selective fading. Nevertheless, single-antenna BS URA schemes suffer from the limited spatial separability among users, degrading system performance as the number of active users grows.


Massive multiple-input-multiple-output (MIMO) \cite{massmimo} is a prominent technology in 5G and beyond communication systems due to the improved capacity, spectral efficiency, and coverage. Many practical URA schemes have been developed for a massive MIMO receiver \cite{agostini,zhang2,decurninge,fengler,fasura2,twc,ozates5,zhang3,zhang4}. Main techniques for exploiting multiple antennas include uncoupled CCS \cite{agostini,zhang2}, employing tensors \cite{decurninge} or pilot sequences \cite{fengler,fasura2,twc} for user separation, and ODMA \cite{ozates5,zhang3,zhang4}. Although massive MIMO significantly improves scalability and energy efficiency in URA, coverage and user fairness remain issues in large-scale fading scenarios.


\subsection{Cell-free massive MIMO and URA}

\begin{figure}
    \centering
    \hspace*{15mm}
     \includegraphics[scale = 0.4]{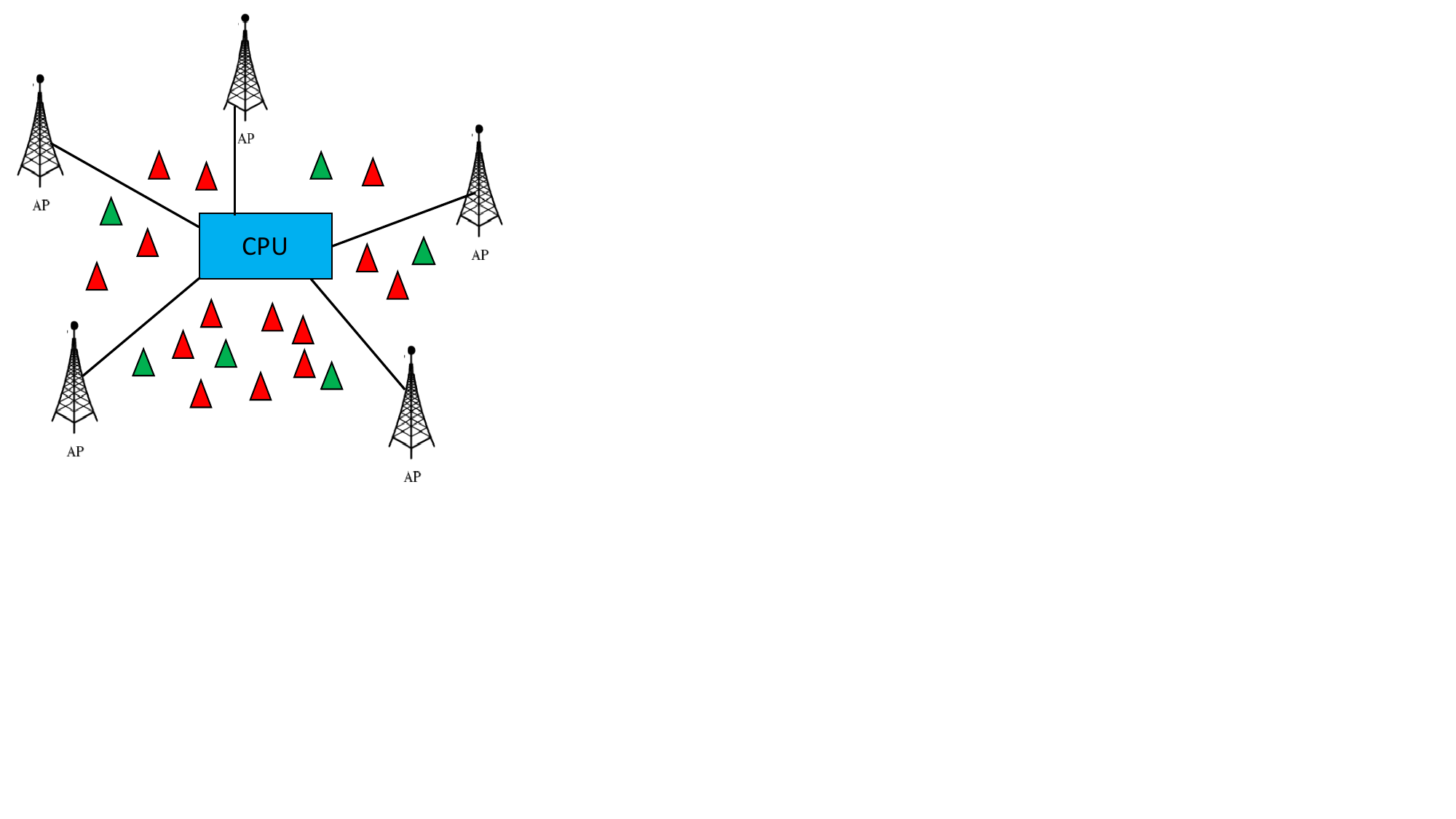}
     \vspace*{-33mm}
    \caption{Illustration of a cell-free massive MIMO system with five access points (APs) and a central processing unit (CPU). Active and inactive users are shown by green and red triangles, respectively.}
    \label{figsystem}
\end{figure}

In cellular MIMO systems, users at the cell edges can experience severe path loss, degrading system performance. In the alternative cell-free massive MIMO paradigm  \cite{bjornson,bjornson2}, rather than a single base station (BS) with many antennas, a large number of access points (APs) with a small number of antennas are deployed (see Fig.\ref{figsystem} for an illustration). The APs are connected to a central processing unit (CPU) and serve users cooperatively. This removes cell boundaries, resulting in better coverage, improved reliability, robustness, and spectral efficiency. However, since all the APs serve all the users, cell-free massive MIMO in its original form is not scalable; namely, the system complexity grows with the number of users. This problem is resolved in \cite{bjornson2} through user-centric clustering, where each AP serves only a subset of users, thereby keeping complexity bounded even as the number of users grows.


Cell-free massive MIMO has also been studied in the URA context \cite{vtc,cefura,cakmak,hu,zhang5,jiang,sun}. In \cite{vtc} and \cite{cefura}, two scalable URA schemes are proposed assuming \textit{Level 2} cooperation, where the transmitted symbols are estimated at the APs, and only the symbol estimates are passed to the CPU. In \cite{vtc}, ODMA is employed for transmission, whereas random spreading, similar to \cite{fasura2}, is used in \cite{cefura}. Instead, the scenario that the APs pass their received signals to the CPU (\textit{Level 4} cooperation) has been studied in \cite{cakmak,hu,zhang5,jiang,sun}. While variations of approximate message passing (AMP) are utilized for receiver processing in \cite{cakmak} and \cite{hu}, an uncoupled CCS-based scheme is proposed in \cite{zhang5}. A multi-level cell-free URA scheme is proposed in \cite{jiang}, in which users are clustered and pass their received signals to the cluster's spatial expansion unit, which acts as a relay between users and the CPU. An asynchronous cell-free URA system is studied in \cite{sun} where a CCS-based scheme employing convolutional neural networks for data detection is proposed.


\subsection{Contributions}

The existing schemes studying URA in a cell-free setup support only a limited number of active users and have limited energy efficiency. For instance, in \cite{cefura}, numerical results are provided up to 150 active users, and our extensive simulations show that the scheme in \cite{cefura} can support up to 450 active users. Motivated by these observations, in this paper, we propose a low-complexity energy-efficient cell-free URA scheme supporting a high number of active users. In this scheme, the transmitted signal vector for each user consists of two parts: a pilot sequence selected from a common non-orthogonal pilot codebook based on a portion of the message bits, and a polar codeword distributed across the data portion of the packet in an ODMA manner. We employ orthogonal matching pursuit (OMP) for joint activity detection and channel estimation and linear minimum mean square error (LMMSE) symbol estimation at the APs. Then, the symbol estimates are passed to the CPU utilizing \textit{Level 2} cooperation. Namely, the symbol estimates belonging to the same user are averaged at the CPU, followed by single-user decoding and successive interference cancellation (SIC) at the end of each iteration. To keep the system scalable, we detect only a subset of the active users at each AP in the \textit{Level 2} cooperation scenario. We also study the centralized scenario in which the APs pass their received signals directly to the CPU, where the active user detection, channel and symbol estimation, single-user decoding, and SIC operations are performed centrally. Note that the channel coefficients are re-estimated using the decoded signals for SIC.

The contributions of the paper can be summarized as follows:

\begin{itemize}
    \item We propose a scalable cell-free URA scheme that leverages ODMA transmission, OMP-based activity detection and channel estimation, LMMSE symbol estimation, single-user decoding, and SIC. Numerical results demonstrate that the proposed scheme outperforms the baseline in \cite{cefura} by up to 9 dB in the single-antenna AP scenario and by up to 5 dB in the multiple-antenna AP scenario, while supporting up to 1800 and 1400 active users, respectively. In addition, the cell-free architecture decreases the PUPE by two orders of magnitude.
    \item We present a finite-blocklength (FBL) performance analysis of the proposed scheme by computing the signal-to-interference-and-noise ratio (SINR) at the input of the polar decoder and applying a normal approximation to obtain a closed-form PUPE expression conditioned on the channel realizations. Numerical examples verify the accuracy of this analysis.
    \item We provide a comprehensive comparison among different cooperation levels. The results reveal that the proposed \textit{Level 2} cooperation either outperforms other cooperation levels or achieves similar performance.
    \item We conduct a detailed complexity analysis and examine the effect of pilot collisions and blockage on system performance.
\end{itemize}


The rest of the paper is organized as follows. We introduce the system model in Section \ref{system} and the proposed scheme in Section \ref{proposed}. We present a performance and complexity analysis of the scheme in Section \ref{perf}, provide a brief analysis of the pilot collisions in Section \ref{results}, and draw conclusions in Section \ref{conclusion}.





\section{System Model} \label{system}

We consider a cell-free scenario with $K_a$ active users out of a massive number of potential users. Each user is equipped with a single antenna and transmits $B$ bits of information to $M$ distributed access points (APs) over a transmission frame of length $n$. Each AP is equipped with $M_r$ antennas and connected to a CPU through a high-capacity fronthaul link. An example of such a system with five APs and a CPU is illustrated in Fig. \ref{figsystem}. The received signal at the $m$-th AP can be written as

\begin{equation}
    \mathbf{Y}_m = \sum\limits_{i =1 }^{K_a} \mathbf{x}_i \mathbf{g}_{i,m} + \mathbf{Z}_m,
\end{equation}

\noindent where $\mathbf{Y}_m \in\mathbb{C}^{n \times M_r}$, $\mathbf{x}_i \in\mathbb{C}^{n \times 1}$ is the transmitted signal of the $i$-th user, $\mathbf{g}_{i,m} \in\mathbb{C}^{1 \times M_r}$ is the vector of channel coefficients of the $i$-th user to the $m$-th AP, and $\mathbf{Z}_m \in\mathbb{C}^{n \times M_r}$ is the circularly symmetric additive white Gaussian noise (AWGN) with independent and identically distributed (i.i.d.) elements from a zero-mean complex normal distribution with variance $\sigma_z^2$. The transmitted symbol sequence of length $n$ satisfies the power constraint $\norm{\mathbf{x}_i}^2 \leq n_pP_p + n_dP_d$, where $n_p$ is the pilot length, $n_d$ is the length of the symbol sequence in the data part, $P_p$ is the average symbol power in the pilot part, and $P_d$ is the average symbol power of the data part. The channel vector $\mathbf{g}_{i,m}$ is defined as \cite{cefura}

\begin{equation}
    \mathbf{g}_{i,m} = \sqrt{\beta_{i,m}} \mathbf{h}_{i,m},
\end{equation}

\noindent where $\mathbf{h}_{i,m}$ is the small-scale fading vector and ${\beta_{i,m}}$ is the large-scale fading coefficient of the $i$-th user to the $m$-th AP. It is assumed that the channel fading coefficients are constant throughout the transmission frame, i.e., quasi-static fading.


Throughout the paper, we consider three cooperation levels: \textit{Level 1}, \textit{Level 2}, and \textit{Level 4} \cite{bjornson}. We propose a cell-free URA scheme employing \textit{Level 2} cooperation, but we also implement \textit{Level 1} and \textit{Level 4} for comparison. In the following, we define their operation and implementation in our scheme.

\begin{itemize}
     \item \textbf{\textit{Level 1} cooperation}: This cooperation level refers to the case where there is no cooperation between the APs. Namely, each AP tries to decode a fixed number of users itself; i.e., each performs the entire decoding procedure independently.
    \item \textbf{\textit{Level 2} cooperation}: In this scenario, activity detection, channel estimation, and symbol estimation are performed at the APs. Then, the symbol estimates are passed to the CPU, where they are simply averaged for symbol combining.
    \item \textbf{\textit{Level 4} cooperation}: In this case, the APs pass their received signals to the CPU, and all decoding operations take place there. This is the highest level of cooperation between the APs.

\end{itemize}

In the \textit{Level 4} cooperation scenario, the concatenated signal $\mathbf{Y} = \left[ \mathbf{Y}_1, \mathbf{Y}_2, \dots \mathbf{Y}_M \right]$ at the CPU can be written as


\begin{equation}
    \mathbf{Y} = \sum\limits_{i =1}^{K_a} \mathbf{x}_i \mathbf{g}_i + \mathbf{Z}
    \label{rxcpu}
\end{equation}

\noindent where $\mathbf{Y} \in\mathbb{C}^{n \times M_\text{tot}}$, $\mathbf{g}_i \in\mathbb{C}^{1 \times M_\text{tot}}$ is the concatenated channel vector of the $i$-th user obtained by combining its channel vectors to different APs, $M_\text{tot} = M_rM$ is the total number of receive antennas, and $\mathbf{Z}$ is the AWGN.

CPU aims to produce a list of the transmitted messages up to a permutation. The system performance is measured by PUPE $P_e = P_{\text{md}} + P_{\text{fa}}$, the summation of the misdetection probability $P_{\text{md}}$ and false alarm probability $P_{\text{fa}}$, defined as

\begin{equation}
  P_{\text{md}} \triangleq  {\mathbb{E}\bigg[ \frac{1}{K_a} \sum\limits_{i \in \mathcal{K}_a} \mathbbm{1}_{\{\mathbf{m}_i \notin \mathcal{L}\}}\bigg]},
\end{equation}

\begin{equation}
   P_{\text{fa}} \triangleq \mathbb{E} \left[\frac{\abs{\mathcal{L} \setminus  \{\mathbf{m}_i: i \in \mathcal{K}_a \}}}{\abs{\mathcal{L}}}\right],
\end{equation}

\noindent where $\mathcal{K}_a$ is the set of active users, $\mathbf{m}_i$ is the message of the ($i$-th) user, $\mathcal{L}$ is the list of decoded messages, $\mathbbm{1}_{\{\ \hspace{-1mm} \cdot \}}$ is the indicator function, and $\abs{\cdot}$ denotes the cardinality of a set.

\section{Proposed Scheme} \label{proposed}

\subsection{Encoding}

In our proposed scheme, each user partitions its $B$-bit message $\mathbf{m}_i \in \{0,1\}^B$ as $\mathbf{m}_i = \left[\mathbf{m}_i^p \hspace{1mm} \mathbf{m}_i^d \right]$, where the leading $B_p$ bits $\mathbf{m}_i^p$ are mapped to a pilot sequence of length $n_p$ that is selected from a common non-orthogonal pilot codebook $\mathbf{A} \in\mathbb{C}^{n_p \times N}$ of $N = 2^{B_p}$ sequences. The pilot part of the received signal at the $m$-th AP and at the CPU (for the centralized operation) can be written as

\begin{equation}
    \mathbf{Y}_{p,m} = \sum\limits_{i =1 }^{K_a} \mathbf{a}_i \mathbf{g}_{i,m} + \mathbf{Z}_{p,m},
\end{equation}

\begin{equation}
    \mathbf{Y}_{p} = \sum\limits_{i =1 }^{K_a} \mathbf{a}_i \mathbf{g}_{i} + \mathbf{Z}_{p},
\end{equation}

\noindent where $\mathbf{Y}_{p,m} \in \mathbb{C}^{n_p \times M_r}$, $\mathbf{Y}_p \in\mathbb{C}^{n_p \times M_\text{tot}}$, $\mathbf{a}_i$ is the selected pilot sequence of the $i$-th user, and $\mathbf{Z}_{p,m}$ and $\mathbf{Z}_{p}$  are the matrices of the first $n_p$ columns of $\mathbf{Z}_{m}$ and $\mathbf{Z}$, respectively. Note that the pilot sequences (columns of $\mathbf{A}$) are normalized to satisfy $\frac{1}{n_p} \norm{\mathbf{a}_i}^2 = P_p$.

\begin{figure}
    \centering
    \hspace*{-5mm}
     \includegraphics[scale = 0.38]{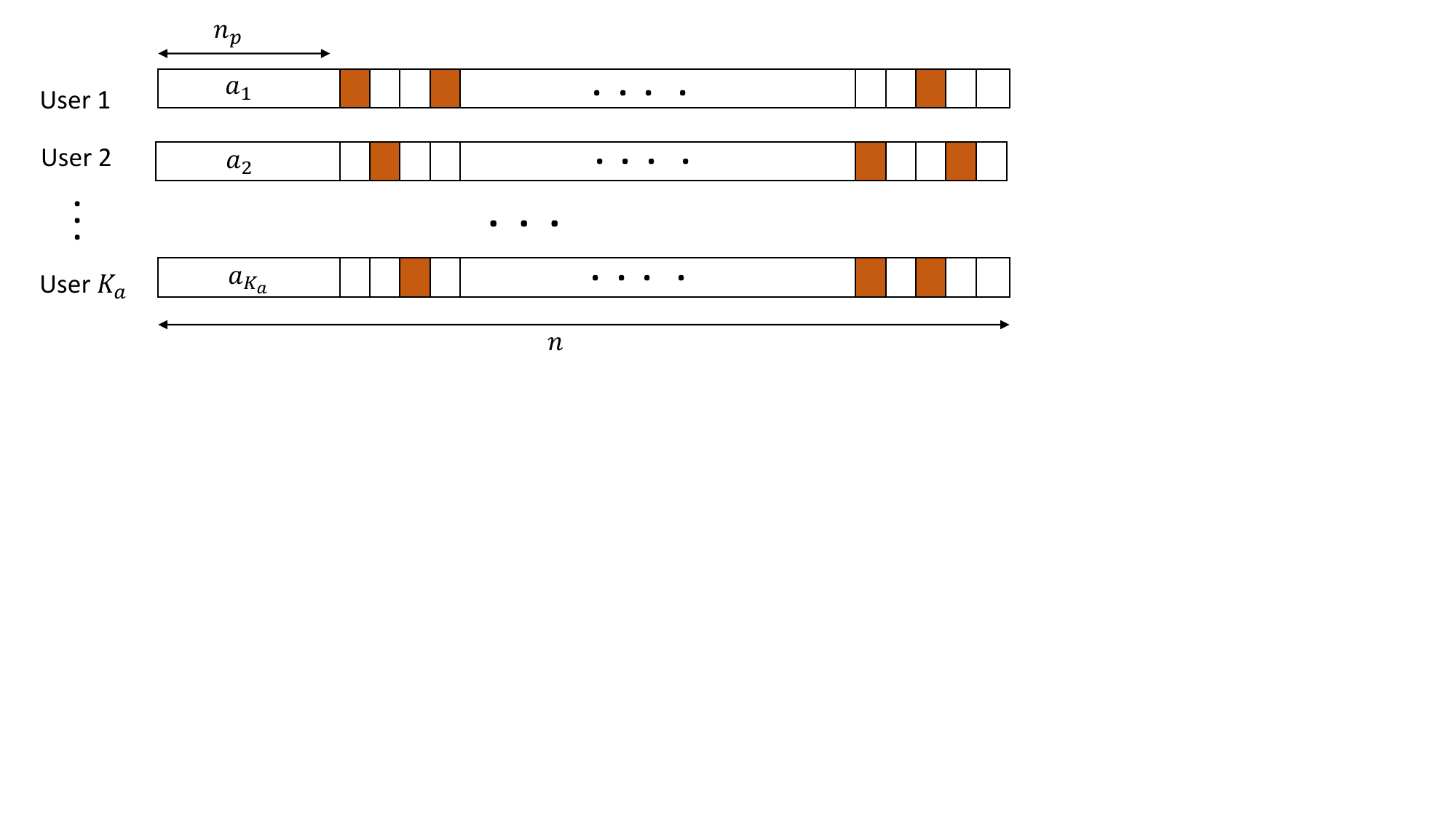}
     \vspace*{-44mm}
    \caption{The ODMA-based transmit signal structure, consisting of the pilot and data parts. Colored boxes indicate the used symbol periods in the data section.}
    \label{encoder}
\end{figure}

The remaining $B -B_p$ bits are appended with $r$ cyclic redundancy check (CRC) bits and encoded by a $(n_c, B-B_p + r)$ polar code, prior to be modulated by quadrature phase shift keying (QPSK) to obtain the symbol sequence $\mathbf{c}_i$ of length $n_d$ with elements from $\{ \sqrt{P_d/2} (\pm 1 \pm j)\}$, satisfying $\frac{1}{n_d} \norm{\mathbf{c}_i}^2 = P_d$. The modulated sequence $\mathbf{c}_i$ is then randomly spread across the $ n-n_p$-length data portion of the frame, according to an ODMA pattern determined by the first $B_p$ bits. The transmission patterns are selected from a pattern codebook $\mathbf{P}' \in \mathbb{F}_2^{(n-n_p) \times N}$ with $n_d$ non-zero elements in each column, indicating the locations of the codeword elements (active indices). The transmit signal structure of the proposed scheme is illustrated in Fig. \ref{encoder}. The received signal $\mathbf{Y}_{d,m} \in\mathbb{C}^{(n - n_p) \times M_r}$ in the data part at the $m$-th AP and at the CPU $\mathbf{Y}_{d} \in\mathbb{C}^{(n - n_p) \times M_{\text{tot}}}$ (for the centralized operation) can be written as


\begin{equation}
  \mathbf{Y}_{d,m} = \sum\limits_{i =1 }^{K_a} \mathbf{P}_i \mathbf{c}_i \mathbf{g}_{i,m} + \mathbf{Z}_{d,m},
\end{equation}

\begin{equation}
   \mathbf{Y}_{d} = \sum\limits_{i =1 }^{K_a} \mathbf{P}_i \mathbf{c}_i \mathbf{g}_{i} + \mathbf{Z}_{d},
\end{equation}

\noindent where $\mathbf{P}_i \in\mathbb{F}_2^{(n-n_p) \times n_d}$ is the pattern matrix of the $i$-th user spreading $\mathbf{c}_i \in\mathbb{C}^{n_d}$ across the $n - n_p$ data symbols, and $\mathbf{Z}_{d,m}$ and $\mathbf{Z}_{d}$ are the matrices consisting of the last $n - n_p$ rows of $\mathbf{Z}_{m}$ and $\mathbf{Z}$, respectively. 

The pattern matrix $\mathbf{P}_i$ is constructed as follows: Letting $\mathbf{P}'_i$ be the length-$n_d$ pattern vector of the $i$-th user denoting the active indices $k \in \{1, 2, \dots n_d'\}$ and assuming that the $j$-th element of $\mathbf{P}'_i$ is $k$, the $(j,k)$-th element of $\mathbf{P}_i$ is set to 1 while the other column elements are zero, where $n_d' = n - n_p$ is the total length of the data part. In other words, each column of $\mathbf{P}_i$ contains one nonzero element, indicating the active index. For example, suppose that the active indices of the first user are on the time instances 3, 4, and 5 in a frame of length 5. Then, $\mathbf{P}'_1 = [3, 4, 5]$, and  $\mathbf{P}_1 = 
\begin{bmatrix}
    0 & 0 & 1 & 0 & 0 \\
    0 & 0 & 0 & 1 & 0 \\
    0 & 0 & 0 & 0 & 1 
\end{bmatrix} ^{\top}$

%

\subsection{Receiver Operation}

At the receiver side, we employ an iterative decoding approach. For the \textit{Level 2} cooperation scenario, at each AP we detect a subset of users (to keep the system scalable) using joint pilot detection and channel estimation, followed by symbol estimation. Then, the estimated symbols are passed to the CPU, where symbol combining and single-user decoding occur. On the other hand, \textit{Level 4} cooperation refers to the centralized operation, where the APs pass their received signals to the CPU, and all the aforementioned decoding operations are performed there without symbol combining. In the following, we explain these processes in detail. Note that at the end of each iteration, the effect of the decoded messages is subtracted by SIC.




\subsubsection{Receiver Operation in Level 2}

\paragraph{Pilot Detection and Channel Estimation} To detect the user messages, the active pilot sequences and their corresponding channel vectors are first estimated. For this purpose, we employ the greedy iterative OMP algorithm \cite{omp}, a well-known method for sparse recovery. That is, we find the pilot sequence with the highest correlation with the received signal at each iteration and subtract its effect by using its projection onto the signal space. First, we calculate the correlation between the candidate pilots and the received signal at the $m$-th AP:


\begin{equation}
    \mathbf{R}_m = \mathbf{A}^H \mathbf{Y}_{\text{OMP},m}^{(k)},
    \label{eqomp}
\end{equation}

\noindent where $\mathbf{R}_m \in\mathbb{C}^{N \times M_r}$, $\mathbf{Y}_{\text{OMP},m}^{(k)}$ is an auxiliary variable that is initialized as $\mathbf{Y}_{\text{OMP},m}^{(1)} = \mathbf{Y}_{p,m}^{(j)}$, where $\mathbf{Y}_{p,m}^{(j)}$ is the residual received pilot signal at the $m$-th AP in the $j$-th decoding iteration (iteration between the APs and CPU), and $\mathbf{Y}_{\text{OMP},m}^{(k)}$ is the residual at the $k$-th OMP iteration. We compute the Euclidean norm of each row $\mathbf{R}_m$ and add the row index corresponding to the maximum Euclidean norm to the output list $\mathcal{\hat{I}}_m$ as the detected pilot. Then, the effect of the detected pilots until the $k$-th OMP iteration is subtracted as


\begin{equation}
 \begin{aligned} 
    \mathbf{\hat{G}}_{m,t}^{(k)} &= \left(\mathbf{A}_{\hat{{I}}_m^{(k)}}^H \mathbf{A}_{\hat{{I}}_m^{(k)}} + \sigma_z^2 \mathbf{I}_{\abs{{\hat{I}}_m^{(k)}}} \right)^{-1} \mathbf{A}_{\hat{{I}}_m^{(k)}}^H \mathbf{Y}_{\text{OMP},m}^{(1)},  \\ 
    \mathbf{Y}_{\text{OMP},m}^{(k + 1)} &=  \mathbf{Y}_{\text{OMP},m}^{(1)}- \mathbf{A}_{\hat{{I}}_m^{(k)}} \mathbf{\hat{G}}_{m,t}^{(k)},
\end{aligned}
 \label{eqsubt}
\end{equation}

\noindent where $\mathbf{A}_{\hat{{I}}_m^{(k)}}$ is the set of the columns of $\mathbf{A}$ for the $m$-th AP specified by ${\hat{{I}}_m^{(k)}}$, $\hat{I}_m^{(k)}$ is the set of detected pilot indices up to the $k$-th OMP iteration, $\mathbf{\hat{G}}_{m,t}^{(k)}$ is the matrix of temporary channel vector estimates corresponding to $\mathbf{A}_{\hat{{I}}_m^{(k)}}$ obtained by LMMSE, and $\mathbf{I}$ denotes the identity matrix. After $K_m$ OMP iterations, the estimated channel vectors become


\begin{equation}
        \mathbf{\hat{G}}_m = \left(\mathbf{\hat{A}}_m^H\mathbf{\hat{A}}_m + \sigma_z^2 \mathbf{I}_{{K_m}} \right)^{-1} \mathbf{\hat{A}}_m^H \mathbf{Y}_{p,m},
    \label{eqest}
\end{equation}

\noindent where $\mathbf{\hat{G}}_m \in \mathbb{C}^{K_m \times M_r}$, and $\mathbf{\hat{A}}_m =  \mathbf{A}_{\hat{{I}}_m^{(K_m)}}\in\mathbb{C}^{n_p \times K_m}$ denotes the detected set of active pilot sequences at the ($m$-th AP). Note that, to simplify the notation, we omit the decoding iteration index, and the equations are presented for the first decoding iteration; however, the same operations can be applied to the residual pilot and data signals in subsequent decoding iterations.

\paragraph{Symbol Estimation} Given the channel vector estimates, using the LMMSE method, the symbol estimates of the $k$-th user at the $m$-th AP are obtained as 

\begin{equation}
    \mathbf{\hat{c}}_{k,m} = \mathbf{\hat{P}}_k^T \mathbf{Y}_{d,m} \left(\mathbf{\hat{G}}_m^H \mathbf{\hat{G}}_m + \frac{\sigma_z^2}{P_d} \mathbf{I}_{M_r} \right)^{-1} \mathbf{\hat{g}}_{k,m}^H,
    \label{eqsymest}
\end{equation}

\noindent where $\mathbf{\hat{g}}_{k,m}$ is the channel vector estimate of the $k$-th user at the $m$-th AP, and $\mathbf{\hat{P}}_k$ is the transmission pattern matrix estimate of the $k$-th user. The symbol estimates are passed to the CPU along with the active pilot indices, as they are required to combine symbol estimates for the same user from different APs. Note that since $\textit{Level 2}$ cooperation is employed, large-scale fading coefficient estimates are not passed to the CPU.

\paragraph{Symbol Combining and Decoding} At the CPU, the symbol estimates from different APs belonging to the same user are combined to obtain the final symbol estimates as follows

\begin{equation}
\mathbf{\hat{c}}_i = \frac{1}{n_i} \sum\limits_{m =1}^{M} \mathbbm{1}_{\{ i\in \mathcal{\hat{I}}_m\}} \hat{\mathbf{C}}_m \left[:,f_m( i ) \right], \quad \forall i\in[N],
  \label{eqcomb}
\end{equation}

\noindent where $n_i\triangleq |\{m:i\in \mathcal{\hat{I}}_m\}|$ is the number of APs that have detected the pilot sequence with index $i$, and $f_m (i)$ is a function mapping the pilot index $i$ to the corresponding column index of $\hat{\mathbf{C}}_m$, that is the matrix of symbol estimates at the $m$-th AP. Note that we link each pilot index estimate to a user, as the number of pilot sequences can be set to make the collision probability small, i.e., the effect of collisions is ignored.


The log-likelihood ratio (LLR) values corresponding to symbol estimates are extracted and fed to a single-user polar decoder employing CRC following the decoding operation. The decoded sequence is added to the output list if the CRC is satisfied. 


\begin{figure*}
    \centering
     \includegraphics[scale = 0.5]{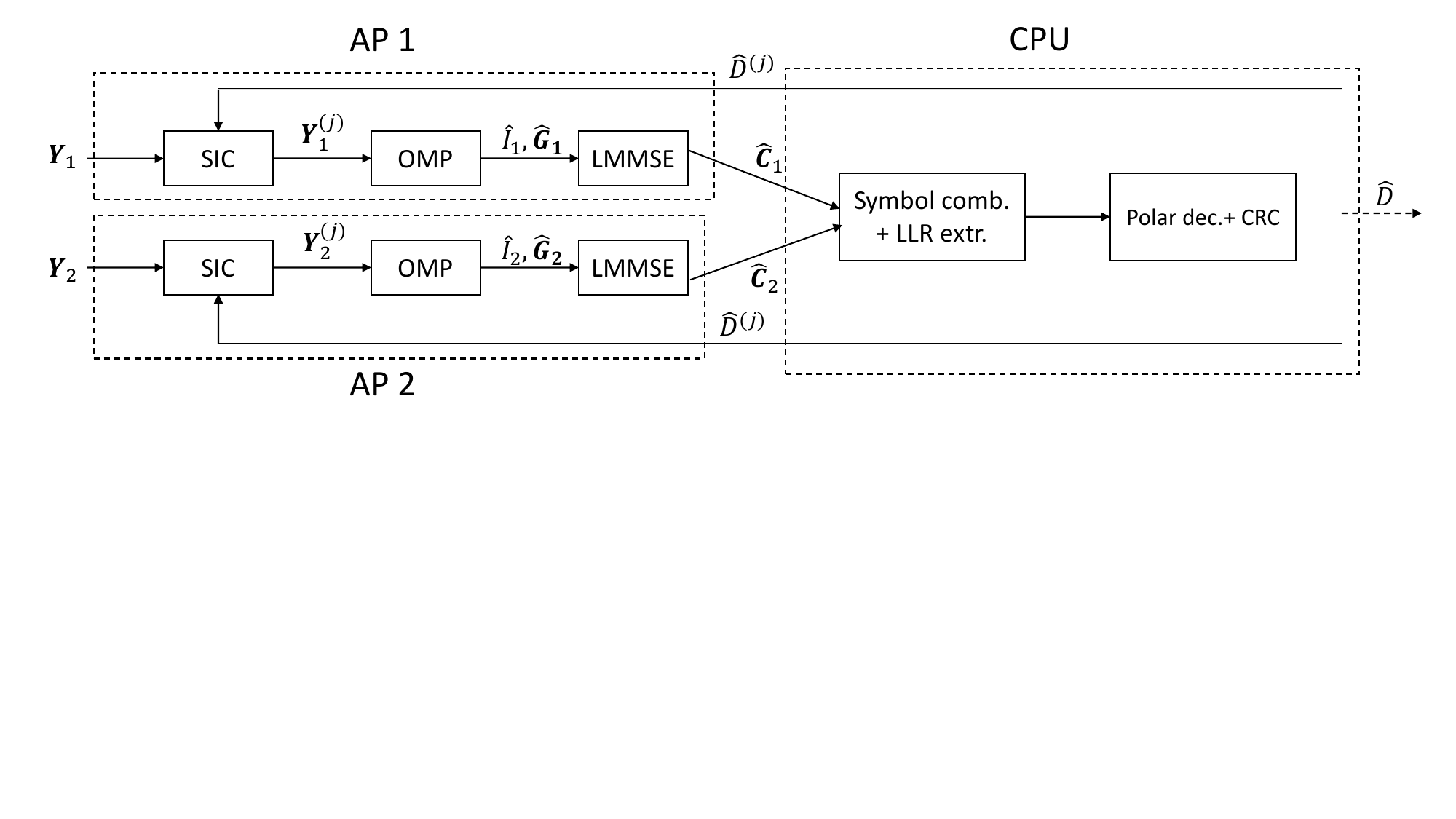}
     \vspace*{-45mm}
    \caption{An illustration of the decoding process of the proposed scheme with two APs, comprising orthogonal matching pursuit (OMP)- based joint pilot and channel estimation, followed by linear minimum mean square error (LMMSE)- based symbol estimation for the APs. The symbol estimates are combined at the CPU utilizing \textit{Level 2} cooperation, and the recovered bits are fed back to APs for successive interference cancellation (SIC). }
    \label{figdecoder}
\end{figure*}

\begin{figure}
    \centering
     \includegraphics[scale = 0.45]{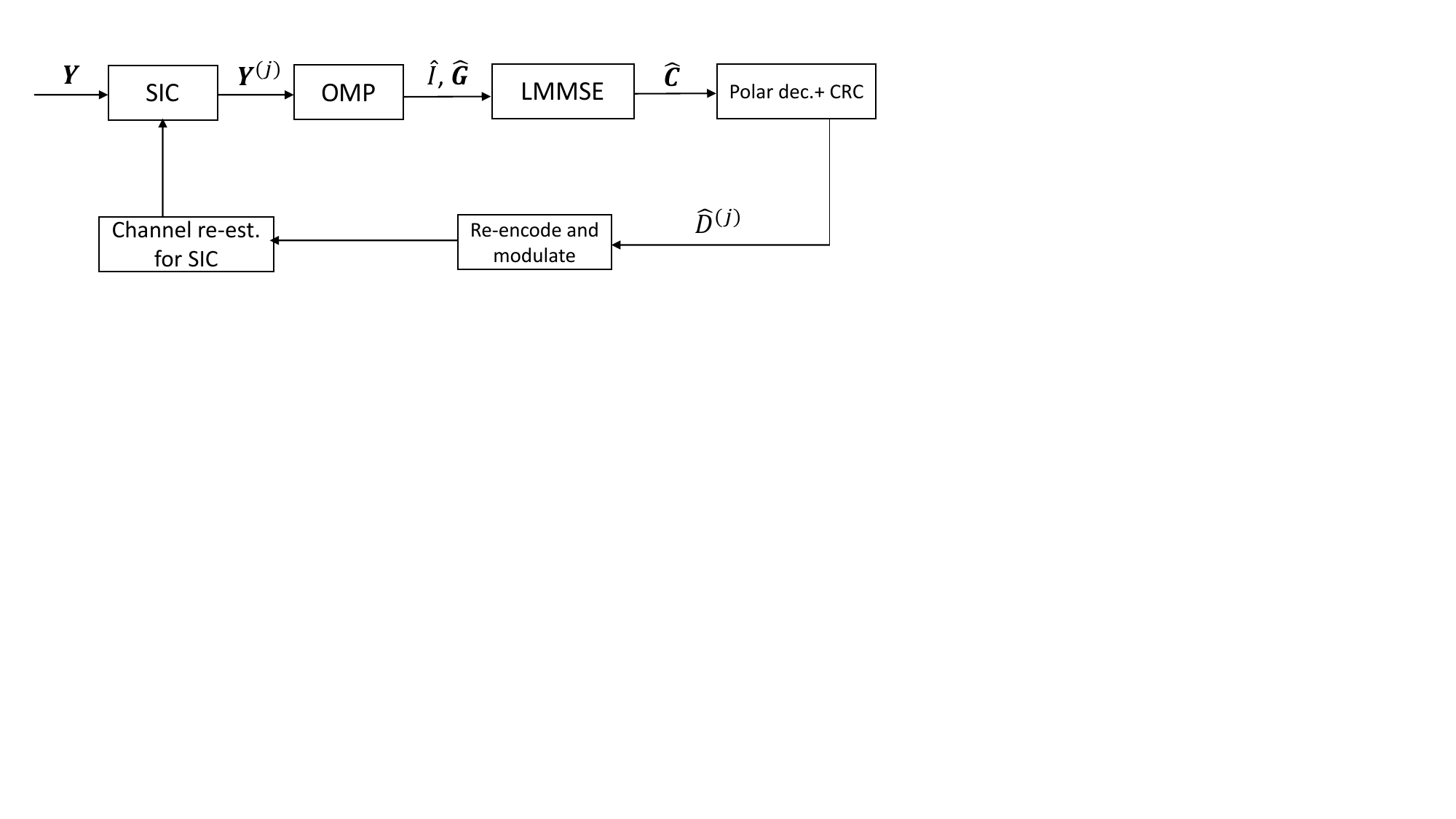}
     \vspace*{-50mm}
    \caption{An illustration of the decoding process of the proposed scheme with centralized operation (\textit{Level 4} cooperation). }
    \label{figdecodercent}
\end{figure}


\paragraph{Successive Interference Cancellation} In URA systems, it is typical to subtract the effect of the decoded user with SIC to cope with high multiuser interference. In our proposed scheme, after the CPU operation, we pass the successfully decoded message bits back to the APs via the fronthaul, where they are re-encoded and modulated prior to SIC. To improve performance, we re-estimate the channel coefficients using the decoded symbols and the pilot sequences, and then utilize the re-estimated channels for SIC, similarly to \cite{twc,fasura2}. Without loss of generality, re-estimation and SIC at the $m$-th AP are performed as follows 

\begin{equation}
    \mathbf{\hat{G}}_{\text{SIC},m} = \left(\mathbf{\hat{X}}_{\hat{\mathcal{D}}^{(j)}}^H\mathbf{\hat{X}}_{\hat{\mathcal{D}}^{(j)}} + \sigma_z^2 \mathbf{I}_\abs{\mathcal{\hat{D}}^ {(j)} } \right )^{-1} \mathbf{\hat{X}}_{\hat{\mathcal{D}}^{(j)}}^H \mathbf{Y}^{(j)}_m,
    \label{eqreest}
\end{equation}

\begin{equation}
        \mathbf{Y}^{(j + 1)}_m = \mathbf{Y}^{(j)}_m - \mathbf{\hat{X}}_{\hat{\mathcal{D}}^{(j)}}  \mathbf{\hat{G}}_{\text{SIC},m} 
        \label{eqsic}
\end{equation}

\noindent where $\mathbf{Y}_m^{(j)}$ is the residual of the received signal at the $m$-th AP, ${\hat{\mathcal{D}}^{(j)}}$ is the set of successfully decoded messages, $\mathbf{\hat{X}}_{\hat{\mathcal{D}}^{(j)}} \in \mathbb{C}^{n \times K_d^j}$ is the matrix including the reconstructed transmitted symbols, where $K_d^j$ is the number of decoded messages in the $j$-th iteration. The residual signal is given to the pilot detection and channel estimation step to start a new iteration.

\subsubsection{Receiver Operation in Level 4}

\paragraph{Pilot Detection and Channel Estimation} In the case of centralized processing, the pilot detection and channel estimation operations can be performed by applying OMP to the pilot signal $\mathbf{Y}_p$ at the CPU. First, the correlation step is performed as

\begin{equation}
    \mathbf{R} = \mathbf{A}^H \mathbf{Y}_{\text{OMP}}^{(k)},
    \label{eqompcent}
\end{equation}

\noindent where $\mathbf{R} \in\mathbb{C}^{N \times M_{\text{tot}}}$, $\mathbf{Y}_{\text{OMP}}$ is an auxiliary variable that is initialized as $\mathbf{Y}_{\text{OMP}}^{(1)} = \mathbf{Y}_{p}^{(j)}$, where $\mathbf{Y}_{p}^{(j)}$ is the residual received pilot signal at the CPU in the $j$-th decoding iteration. Then, the effect of the detected pilots until the $k$-th OMP iteration is subtracted as

\begin{equation}
 \begin{aligned} 
    \mathbf{\hat{G}}_{t}^{(k)} &= \left(\mathbf{A}_{\hat{{I}}^{(k)}}^H \mathbf{A}_{\hat{{I}}^{(k)}} + \sigma_z^2 \mathbf{I}_{\abs{{\hat{I}}^{(k)}}} \right)^{-1} \mathbf{A}_{\hat{{I}}^{(k)}}^H \mathbf{Y}_{\text{OMP}}^{(1)},  \\ 
    \mathbf{Y}_{\text{OMP}}^{(k + 1)} &=  \mathbf{Y}_{\text{OMP}}^{(1)}- \mathbf{A}_{\hat{{I}}^{(k)}} \mathbf{\hat{G}}_{t}^{(k)},
\end{aligned}
 \label{eqsubtcent}
\end{equation}

\noindent where $\mathbf{A}_{\hat{{I}}^{(k)}}$ is the set of the columns of $\mathbf{A}$ specified by ${\hat{{I}}^{(k)}}$, $\hat{I}^{(k)}$ is the set of detected pilot indices up to the $k$-th OMP iteration, and $\mathbf{\hat{G}}_{t}^{(k)}$ is the matrix of temporary channel vector estimates corresponding to $\mathbf{A}_{\hat{{I}}^{(k)}}$ obtained by LMMSE. After $K_a$ OMP iterations, the estimated channel vectors become 

\begin{equation}
        \mathbf{\hat{G}} = \left(\mathbf{\hat{A}}^H\mathbf{\hat{A}} + \sigma_z^2 \mathbf{I}_{{K_a}} \right)^{-1} \mathbf{\hat{A}}^H \mathbf{Y}_{p},
    \label{eqestcent}
\end{equation}

\noindent where $\mathbf{\hat{G}} \in \mathbb{C}^{K_a \times M_{\text{tot}}}$, and $\mathbf{\hat{A}} \in\mathbb{C}^{n_p \times K_a}$ denotes the detected set of active pilot sequences. Note that $K_a$ is assumed to be known at the CPU.

\paragraph{Symbol Estimation}

When centralized processing is employed, symbol estimation occurs on the CPU, jointly for all users. In this case, the symbol estimates of the $k$-th user can be calculated as

\begin{equation}
  \hat{\mathbf{c}}_k = \mathbf{\hat{P}}_k^T \mathbf{Y}_{d} \left(\mathbf{\hat{G}}^H \mathbf{\hat{G}}+ \frac{\sigma_z^2}{P_d} \mathbf{I}_{M_{\text{tot}}} \right)^{-1} \mathbf{\hat{g}_k}^H ,
    \label{eqsymestcent2}
\end{equation}

\noindent where $\mathbf{\hat{g}}_{k}$ is the concatenated channel vector estimate of the $k$-th user. Then, the LLRs of all users are extracted from the symbol estimates, and single-user polar decoding and CRC are performed similarly to \textit{Level 2} scenario.


\paragraph{Successive Interference Cancellation} 

In this case, the successfully recovered message bits are directly re-encoded and modulated at the CPU, and their effects are subtracted from the received signal $\mathbf{Y}$ to start a new iteration. The re-estimation of the channel vectors and SIC in the centralized scenario is performed as


\begin{equation}
    \mathbf{\hat{G}}_{\text{SIC}} = \left(\mathbf{\hat{X}}_{\hat{\mathcal{D}}^{(j)}}^H\mathbf{\hat{X}}_{\hat{\mathcal{D}}^{(j)}} + \sigma_z^2 \mathbf{I}_\abs{\mathcal{\hat{D}}^ {(j)} } \right )^{-1} \mathbf{\hat{X}}_{\hat{\mathcal{D}}^{(j)}}^H \mathbf{Y}^{(j)},
    \label{eqreestcent}
\end{equation}

\begin{equation}
        \mathbf{Y}^{(j + 1)} = \mathbf{Y}^{(j)} - \mathbf{\hat{X}}_{\hat{\mathcal{D}}^{(j)}}  \mathbf{\hat{G}}_{\text{SIC}}
        \label{eqsic2cent}
\end{equation}

\noindent where $\mathbf{\hat{X}}_{\hat{\mathcal{D}}^{(j)}} \in  $ is the matrix including the reconstructed transmitted symbols, and $\mathbf{Y}^{(j)}$ is the residual of the received signal.

The decoding iterations continue until either no new user message can be successfully decoded in the current iteration or $n_{\text{max}}$ iterations have been reached. The decoding procedures of the proposed scheme for \textit{Level 2} cooperation and for the centralized scenario are illustrated in Fig. \ref{figdecoder} and Fig. \ref{figdecodercent}, respectively. Moreover, pseudocode for the receiver operation is given in Algorithm \ref{alg} for the \textit{Level 2} scenario, and in Algorithm \ref{algcent} for the centralized scenario where $\mathbf{R} \in\mathbb{C}^{N \times M_{\text{tot}}}$, $\mathbf{Y}_{\text{OMP}}$ is an auxiliary variable that is initialized as $\mathbf{Y}_{\text{OMP}}^{(1)} = \mathbf{Y}_{p}^{(j)}$, where $\mathbf{Y}_{p}^{(j)}$ is the residual received pilot signal at the CPU in the $j$-th decoding iteration for the centralized operation.

\begin{algorithm}[t]
\caption{Receiver operation of the proposed cell-free URA scheme with \textit{Level 2} cooperation}\label{alg}
\begin{algorithmic}[1]

\State \textbf{Input}: $\mathbf{Y}_{p,m}$, $\mathbf{Y}_{d,m}$ for $m = 1,2, \dots M$, $\mathbf{A}$, $n_{\text{max}}$,

\For{$j=1,2, \ldots n_{\text{max}} $}

\State \textbf{Operation at the APs}:

\State \textbf{Activity detection and channel estimation}:
\For{$k=1,2, \ldots K_m $}
\State $ \mathbf{R}_m = \mathbf{A}^H \mathbf{Y}_{\text{OMP},m}^{(k)} $.
\State Get the decision metrics by the Euclidean norm of rows of $\mathbf{R}_m$, take the index with the maximum decision metric, and add it to ${\hat{I}_m}$.
\State Subtract the effect of the detected indices by (\ref{eqsubt}).
\EndFor
\State Estimate the channel vectors by (\ref{eqest}).
\State \textbf{Symbol Estimation}:
\State Estimate the transmitted symbols by (\ref{eqsymest}).

\State Pass symbol estimates and active pilot indices to CPU.

\State \textbf{Operation at the CPU}:
\State Combine the symbol estimates by (\ref{eqcomb}).

\State Set $\mathcal{\hat{D}}^ {(j)} = \emptyset$.

\For{$i=1,2,\ldots, N$}
\State Pass $\mathbf{\hat{c}}_i$ to the polar decoder $\rightarrow$ $\mathbf{\hat{m}}_i$.
\State Cyclic Redundancy Check.
\EndFor
\If {$  \abs{\mathcal{\hat{D}}^ {(j)} } = 0$}
\State Algorithm termination.
\EndIf
\State Pass the successfully decoded bits to APs.
\For{i in $ \mathcal{\hat{D}}^ {(j)}$}
\State Re-encode and modulate $\mathbf{\hat{m}}_i$ 
\EndFor
\State \textbf{SIC}:
\State Interference cancellation by (\ref{eqreest}) and (\ref{eqsic}).

\EndFor
\State $\textbf{Output}$: List of the decoded messages

\end{algorithmic}
\end{algorithm}

\begin{algorithm}[t]
\caption{Receiver operation of the proposed cell-free URA scheme with centralized operation (\textit{Level 4} cooperation)}\label{algcent}
\begin{algorithmic}[1]

\State \textbf{Input}: $\mathbf{Y}_{p}$, $\mathbf{Y}_{d}$, $\mathbf{A}$, $n_{\text{max}}$,

\For{$j=1,2, \ldots n_{\text{max}} $}

\State \textbf{Activity detection and channel estimation}:
\For{$k=1,2, \ldots K_a $}
\State $ \mathbf{R} = \mathbf{A}^H \mathbf{Y}_{\text{OMP}}^{(k)} $.
\State Compute the decision metrics as the Euclidean norms of the rows of $\mathbf{R}$, select the index with the maximum metric, and add it to ${\hat{I}}$.
\State Subtract the effect of the detected indices.
\EndFor
\State Estimate the channel vectors.

\State Estimate the transmitted symbols by  (\ref{eqsymestcent2}).

\State Set $\mathcal{\hat{D}}^ {(j)} = \emptyset$.

\For{$i=1,2,\ldots, N$}
\State Pass $\mathbf{\hat{c}}_i$ to the polar decoder $\rightarrow$ $\mathbf{\hat{m}}_i$.
\State Cyclic Redundancy Check.
\EndFor
\If {$  \abs{\mathcal{\hat{D}}^ {(j)} } = 0$}
\State Algorithm termination.
\EndIf
\For{i in $ \mathcal{\hat{D}}^ {(j)}$}
\State Re-encode and modulate $\mathbf{\hat{m}}_i$. 
\EndFor
\State \textbf{SIC}:
\State Interference cancellation by (\ref{eqreestcent}) and (\ref{eqsic2cent}).

\EndFor
\State $\textbf{Output}$: List of the decoded messages

\end{algorithmic}
\end{algorithm}

\section{Performance Analysis} \label{perf}

\subsection{Error Probability}

In this section, we present an error-probability analysis of the proposed scheme in the finite blocklength (FBL) regime, similarly to that in \cite{twc}. For this purpose, we compute the SINR at the output of the symbol-estimation step (i.e., the input to the polar decoder) for each user, and then employ a normal approximation to estimate the error probability. In the symbol estimation step, users are separated; hence, each user's symbol estimate can be treated as the output of a virtual single-user channel with a signal-to-noise ratio (SNR) equal to their effective SINR. Therefore, the finite blocklength analysis based on normal approximations can be applied. To simplify the analysis, we assume that there are no collisions, the pilot sequences and transmission patterns are perfectly detected, and the channel vectors are known at the receiver. 

\subsubsection{Error Probability in \textit{Level 2}}

In the Level 2 scenario, at each AP, we assume that $K_m$ users with the strongest channel vectors (i.e., those with the highest Euclidean norm) are selected as the detected users. Then, the symbol estimate of the $k$-th user at the $m$-th AP and its final estimate at the CPU can be written as

\begin{equation}
    \mathbf{\hat{c}}_{k,m} = \mathbf{{P}}_k^T \mathbf{Y}_{d,m} \left(\mathbf{{G}}_m^H \mathbf{{G}}_m + \frac{\sigma_z^2}{P_d} \mathbf{I}_{M_r} \right)^{-1} \mathbf{{g}}_{k,m}^H,
\end{equation}

\begin{equation}
\mathbf{\hat{c}}_k = \frac{1}{n_k} \sum\limits_{m =1}^{M} \mathbbm{1}_{\{ k\in \mathcal{\hat{I}}_m\}} \hat{\mathbf{C}}_m \left[:,f_m( k ) \right], \quad \forall k\in[N],
  \label{eqcomb2}
\end{equation}

\noindent where $\mathbf{\hat{c}}_{k,m} \in\mathbb{C}^{n_d \times 1}$, and $\mathbf{G}_m$ is the matrix of the channel vectors of the users detected at the $m$-th AP. Defining  $ W_m \triangleq \left(\mathbf{{G}}_m^H \mathbf{{G}}_m + \frac{\sigma^2}{P_d} \mathbf{I}_{M_r} \right)$ as $\mathbf{W}_m$, the symbol estimate can be rewritten as


\begin{flalign}
    \mathbf{\hat{c}}_k =& \frac{1}{n_k} \sum\limits_{m =1}^{M} \mathbbm{1}_{\{ k\in \mathcal{\hat{I}}_m\}} \mathbf{{P}}_k^T \mathbf{Y}_{d,m} \mathbf{W}_m^{-1} \mathbf{{g}}_{k,m}^H  \\& \hspace*{-4mm}= \frac{1}{n_k} \sum\limits_{m =1}^{M} \mathbbm{1}_{\{ k\in \mathcal{{I}}_m\}} \mathbf{{P}}_k^T \left(  \sum\limits_{i =1 }^{K_a} \mathbf{P}_i \mathbf{c}_i \mathbf{g}_{i,m}+ \mathbf{Z}_m\right) \mathbf{W}_m^{-1} \mathbf{{g}}_{k,m}^H \notag \\& \hspace*{-4mm}= \frac{1}{n_k} \sum\limits_{m =1}^{M} \mathbbm{1}_{\{ k\in \mathcal{{I}}_m\}} \Bigl(\mathbf{{P}}_k^T \mathbf{{P}}_k \mathbf{{c}}_k\mathbf{g}_{k,m} \mathbf{W}_m^{-1} \mathbf{g}_{k,m}^H + \notag  \\&  \sum\limits_{i =1, i \neq k}^{K_a} \mathbf{{P}}_k^T \mathbf{{P}}_i \mathbf{{c}}_i\mathbf{g}_{i,m} \mathbf{W}_m^{-1} \mathbf{g}_{k,m}^H + \mathbf{{P}}_k^T \mathbf{Z}_m \mathbf{W}_m^{-1} \notag \mathbf{g}_{k,m}^H  \Bigr),
\end{flalign}

\noindent where the first term is the desired signal, the second term is the interference, and the last term is the noise. Note that $\mathbf{P}_k^T \mathbf{P}_k = \mathbf{I}_{n_d}$. Then, the SINR of the $k$-th user can be calculated as

\begin{equation}
    \gamma_k = \frac{P_S}{P_I + P_N}.
    \label{eqsinr3}
\end{equation}

The three terms $P_S, P_I, P_N$ in (\ref{eqsinr3}) are calculated as

\begin{equation}
    P_S = \scalebox{0.80}{$\frac{1}{n_k^2} n_dP_d \sum\limits_{m' =1}^{n_k} \sum\limits_{m'' =1}^{n_k}  \mathbf{g}_{k,m'} (\mathbf{W}_{m'}^{-1})^H \mathbf{g}_{k,m'}^H \mathbf{g}_{k,m''} \mathbf{W}_{m''}^{-1} \mathbf{g}_{k,m''}^H$},
\end{equation}

\begin{equation}
    P_I = \scalebox{0.80}{$\frac{1}{n_k^2} P_d \sum\limits_{i =1, i \neq k}^{K_a}  tr\{\mathbf{{P}}_i^T \mathbf{{P}}_k  \mathbf{{P}}_k^T \mathbf{{P}}_i \}  \norm{\sum\limits_{m''=1}^{n_k} \mathbf{g}_{i,m''} \mathbf{W}_{m''}^{-1} \mathbf{g}_{k,m''}^H}^2 $},
\end{equation}

\begin{equation}
\hspace{-30mm}
    P_N =  \frac{1}{n_k^2} n_d \sigma_z^2 \sum\limits_{m' =1}^{n_k}\norm{\mathbf{W}_{m'}^{-1} \mathbf{g}_{k,m'}^H}^2.
\end{equation}

\begin{IEEEproof}
        See the Appendix.
\end{IEEEproof}

\subsubsection{Error Probability in \textit{Level 4}}
    
We also conduct the same procedure for SINR calculation in the centralized scenario. In the first step, the symbol estimate of the $k$-th user at the output of LMMSE can be written as

\begin{equation}
    \hat{\mathbf{c}}_k = \mathbf{P}_k^T \mathbf{Y}_{d} \left(\mathbf{{G}}^H \mathbf{{G}}+ \frac{\sigma_z^2}{P_d} \mathbf{I}_{M_{\text{tot}}} \right)^{-1} \mathbf{{g}_k}^H ,
    \label{eqsymestcent3}
\end{equation}

\noindent where $\hat{\mathbf{c}}_k\in\mathbb{C}^{n_d \times 1} $. Denoting $\left(\mathbf{{G}}^H \mathbf{{G}}+ \frac{\sigma^2}{P_d} \mathbf{I}_{M_{\text{tot}}} \right)$ as $\textbf{W}$, $\hat{\mathbf{c}}_k $ can be rewritten as

\begin{flalign}
      \hat{\mathbf{c}}_k &= \mathbf{P}_k^T  \left(\sum\limits_{i =1 }^{K_a} \mathbf{P}_k \mathbf{c}_k \mathbf{g}_k + \mathbf{Z}_d\right) \mathbf{W}^{-1} \mathbf{{g}_k}^H \\&= \mathbf{P}_k^T \mathbf{P}_k \mathbf{c}_k \mathbf{g}_k  \mathbf{W}^{-1} \mathbf{{g}_k}^H + \left(\sum\limits_{j=1, j \neq k }^{K_a} \mathbf{P}_k^T \mathbf{P}_j \mathbf{c}_j \mathbf{g}_j \mathbf{W}^{-1} \mathbf{{g}_k}^H  \right) \notag \\&+ \mathbf{P}_k^T \mathbf{Z}_d \mathbf{W}^{-1} \mathbf{{g}_k}^H, \notag
\end{flalign}

\noindent where the first term is the desired signal, the second term is the interference, and the third term is the noise. Then, the SINR of the $k$-th user in the centralized scenario can be calculated as

\begin{equation}
    \gamma_{k,c} = \frac{P_{S,c}}{P_{I,c} + P_{N,c}},
    \label{eqsinr2}
\end{equation}

\noindent where $P_{S,c}$ is the power of the desired signal, $P_{I,c}$ is the power of the multiuser interference, and $P_{N,c}$ is the power of the Gaussian noise after MMSE. The three terms in (\ref{eqsinr2}) can be calculated as

\begin{equation}
   \hspace{-40mm}  P_{S,c} = n_dP_d \norm{\mathbf{g}_k  \mathbf{W}^{-1} \mathbf{{g}_k}^H}^2,
\end{equation}

\begin{equation}
    P_{I,c} = \scalebox{0.8}{$P_d \mathbf{g}_k \mathbf{W}^{-1} \left(\sum\limits_{j=1, j \neq k  }^{K_a} tr\{\mathbf{P}_j^T \mathbf{P}_k  \mathbf{P}_k^T \mathbf{P}_j\} \mathbf{{g}_j}^H \mathbf{{g}_j}\right) \mathbf{W}^{-1} \mathbf{g}_k^H $},
\end{equation}

\begin{equation}
  \hspace{-40mm}   P_{N,c} = n_d \sigma_z^2 \norm{\mathbf{W}^{-1} \mathbf{{g}_k}^H}^2. 
\end{equation}

\begin{IEEEproof}
        See the Appendix.
\end{IEEEproof}


For a channel code with blocklength $n_c$, the achievable rate to get an error probability of $p_e$ over the transmission on an AWGN channel is given by the normal approximation \cite{poor}

\begin{equation}
    R \approx C - \sqrt{\frac{V_{dis}}{n_c}}Q^{-1}(p_e),
    \label{eqnormal}
\end{equation}

\noindent where $Q(.)$ is the standard Q-function, $R$ is the code rate, $C$ is the channel capacity, and $V_{dis}$ is the channel dispersion. Moreover, \cite{yang} shows that a quasi-static channel is conditionally ergodic given the channel gain. Then, for a given set of channel coefficients, the SINR can be treated as known, and the channel capacity and dispersion can be calculated as

\begin{equation}
\begin{aligned}
         C &=  \log_2(1+\gamma_k), \hspace{2mm} V_{dis} =  \frac{\gamma_k(\gamma_k+2)}{(\gamma_k+1)^2} \log_2^2 (e).
\end{aligned}
    \label{eqcap}
\end{equation}

Conditioned on the channel gains and given the channel capacity and dispersion in (\ref{eqcap}), the block error probability can be calculated as

\begin{equation}
    p_e \approx Q \left(\frac{C-R}{\sqrt{V_{dis}/n_c}}\right).
\end{equation}

Then, the error probability across the users can be calculated by averaging over different channel realizations as

\begin{equation}
    p_{e,avg} \approx \mathbb{E}_H\left[ Q \left(\frac{C-R}{\sqrt{V_{dis}/n_c}}\right)  \right],
    \label{eqerrorfinal}
\end{equation}

\noindent where the expectation is taken over the channel vectors. The expectation in (\ref{eqerrorfinal}) can be evaluated by Monte Carlo simulations to obtain the PUPE.


\subsection{Complexity}

In this subsection, a computational complexity analysis of the decoding operations in the proposed scheme, in terms of the number of multiplications, is provided. For the \textit{Level 2} cooperation scenario, we present the complexity orders for the operation at each AP, as the operations in different APs are identical. 

For the \textit{Level 2} cooperation, the correlation step in (\ref{eqomp}) in joint pilot detection and channel estimation has a complexity of $\mathcal{O} (Nn_pM_r)$, and the complexity of the dominant matrix multiplication and the matrix inversion in (\ref{eqsubt}) are $\mathcal{O} (K_m^2n_p)$ and $\mathcal{O} (K_m^3)$, respectively, where $\mathcal{O}(\cdot)$ denotes the standard big-O notation. The complexity of matrix inversion in symbol estimation in (\ref{eqsymest}) is $\mathcal{O} (M_r^3)$, and the dominant matrix multiplication has a complexity of $\mathcal{O} (n_dM_rK_m)$. Finally, SIC has a complexity of $\mathcal{O} ((n_d + n_p)K_mM_r)$, which is determined by the matrix multiplication in (\ref{eqsic}). For practical system parameters, the complexity of joint pilot detection and channel estimation is dominated by the matrix multiplication in (\ref{eqomp}) with a complexity of $\mathcal{O} (Nn_pM_r)$, and the dominant complexity of symbol estimation is determined by the matrix multiplication in (\ref{eqsymest}) with $\mathcal{O} (n_dM_rK_m)$.





In the centralized operation, the decoding operations are the same as the \textit{Level 2} scenario, except that there is no symbol combining. Therefore, the complexity of the decoding operations can be obtained by replacing $K_m$ with $K_a$ and $M_r$ with $M_\text{tot}$, while the complexity of channel re-estimation for SIC is $\mathcal{O} (K_a^2n)$.


Let us compare the complexity of the proposed scheme with that of \cite{cefura}, which also employs \textit{Level 2} cooperation. In that scheme, the complexity of joint pilot detection and channel estimation is similar to ours. However, the matrix inversion in symbol estimation has complexity $\mathcal{O}(L^3M_r^3n_d)$, which dominates the complexity of the symbol estimation, where $L$ is the length of the spreading sequence in \cite{cefura}. For practical system parameters, it is much higher than the complexity of the symbol estimation in our method since $L^3 M_r^2 \gg K_m$. Moreover, modifying the scheme in \cite{cefura} to include an SIC block results in a complexity of $\mathcal{O} (nK_mM_r)$, which is slightly higher than that of the proposed scheme. We provide a complexity comparison of our scheme with \textit{Level 2} cooperation and centralized operation and with the scheme in \cite{cefura}, called CEFURA, in Table \ref{tablecomp}, where JADCE stands for joint activity detection and channel estimation.

\subsection{Collisions}

In massive random-access setups, it is not feasible to assign distinct pilots to users due to the huge number of potential users. Therefore, active users often select a pilot sequence from a common codebook based on part of their message bits, and a pilot collision occurs when multiple users select the same pilot sequence. Collisions can degrade system performance by making user separation harder. They are especially important for \textit{Level 2} cell-free operation, as each pilot sequence is treated as a user in the symbol-combining step; hence, only one user can be decoded per iteration. Thus, it is important to make the pilot collision probability negligible by having a sufficiently large pilot codebook. A detailed analysis of the pilot collisions on the system performance for URA with a massive MIMO receiver is provided in \cite{twc}.

\begin{table}
\small
\centering
\caption{Comparison of the complexity orders.}
\begin{adjustbox}{width=1\columnwidth,center}
\begin{tabular}{| c | c | c | c |} 
 \hline
 URA scheme & JADCE & Symbol est. & SIC \\ [0.5ex] 
 \hline\hline
 Proposed, \textit{Level 2}  & $\mathcal{O} (Nn_pM_r)$ & $\mathcal{O} (n_dM_rK_m)$ & $\mathcal{O} ((n_d + n_p)K_mM_r)$   \\ 
 \hline
  Proposed, centralized   & $\mathcal{O} (Nn_pM_\text{tot})$  & $\mathcal{O} (n_dM_\text{tot}K_a)$ & $\mathcal{O} (K_a^2n)$  \\
 \hline
 CEFURA \cite{cefura} & $\mathcal{O} (Nn_pM_r)$ & $\mathcal{O} (L^3M_r^3n_d)$ & $\mathcal{O} (nK_mM_r)$\footnotemark   \\
 \hline
\end{tabular}
\end{adjustbox}
\label{tablecomp}
\normalsize
\end{table}

\footnotetext{Note that the original version of the scheme in \cite{cefura} does not include SIC. However, to make a more direct comparison, we have modified it by adding an SIC block and present the corresponding results in Section \ref{results}.}

\section{Numerical Results} \label{results}

In this section, we assess the performance of the proposed cell-free URA scheme. We set $n = 3200$, $B = 100$, $K_m = 7$, $B_p = 15$, $M = 100$, $M_r = 1$, and $n_p = 800$ unless otherwise specified. We assume that the APs are randomly distributed over a $D \times D$ $m^2$ area according to a binomial point process, with $D = 550$ unless otherwise specified. The elements of the pilot codebook are drawn from a zero-mean standard normal distribution. A successive cancellation list decoding (SCLD) algorithm is employed for polar decoding, with a polar code length of 1024, a CRC length of 16, and an SCLD list size of 8.


We model path loss using the urban micro-cell propagation model in \cite{stand} at a center frequency of 2 GHz, as in \cite{cefura, bjornson}. Then, the large-scale fading coefficient is defined as 

\begin{equation}
    \beta_{i,m} [dB] = -30.5 - 36.7\hspace{1mm}\text{log}_{10}d_{im} + F_{im},
\end{equation}

\noindent where $d_{im}$ is the distance between the $i$-th user and the $m$-th AP, and $F_{im}$ is the shadow fading. The shadow fading model in \cite{cefura} is utilized, namely, $F_{im} \sim \mathcal{N} (0,16)$, and it is correlated from an AP to different users as \cite{stand}

                \begin{flalign}
                \mathbb{E} [F_{im}F_{kj}] =
            \begin{cases}
                           16 \times 2^{-d'_{ik}/9} & \text{if $m=j$} \\ 
                             0, & \text{if $m \neq j$}
                        \end{cases} 
                \end{flalign}
                
\noindent where $d'_{ik}$ is the distance between the $i$-th and $k$-th user. The small-scale fading coefficients are generated by assuming a uniform linear array (ULA) at the APs with half-wavelength antenna spacing and using the spatial correlation matrix provided in (2.23) in \cite{massmimo}. (See Sections 2.2 and 2.6 in \cite{massmimo} for further details on correlated fading).

\begin{figure} 
    \centering
    \includegraphics[width=1\linewidth]{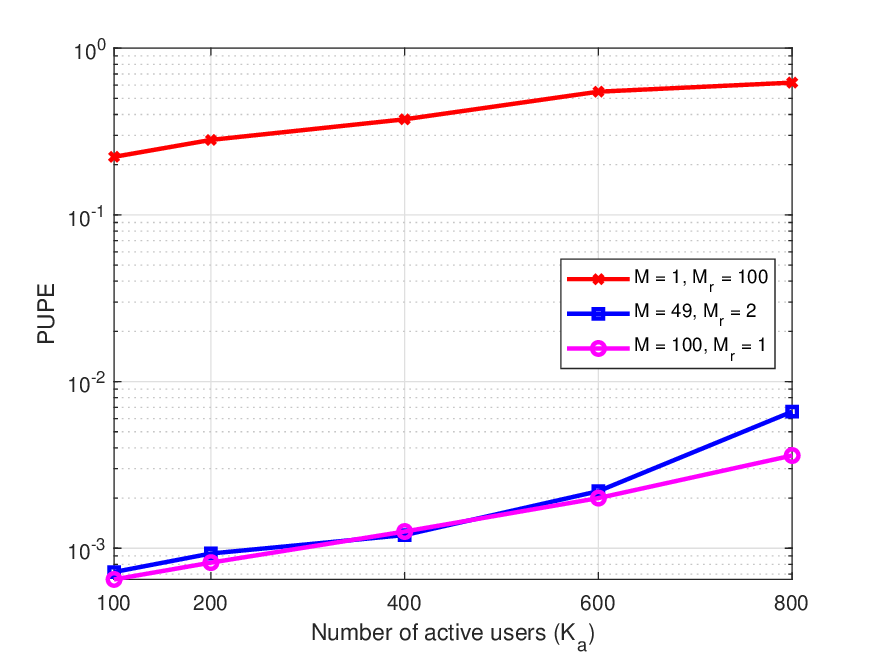}
    \caption{Comparison of PUPE versus the number of active users for different AP configurations.}
    \label{figdist}
\end{figure}

We study the effect of distributing the antennas by evaluating the PUPE for different numbers of active users in Fig. \ref{figdist} for the co-located scenario ($M = 1, M_r = 100$) and two distributed configurations, namely, $M = 100, M_r = 1$ and $M = 49, M_r = 2$.\footnote{Note that we assume a square grid, hence, the number of APs should be square of a number. Therefore, when $M_r = 2$, $M$ is set to 49, as this makes the total number of antennas nearly equal.}  We set the average symbol power to 10 mW and $\sigma^2 = -84$ dBm. It can be observed that distributing the antennas across the geographic area via APs considerably improves system performance; namely, the PUPE in the distributed scenario is two orders of magnitude lower than that of the co-located scenario. In addition, the single-antenna AP scenario outperforms the $M = 2, M_r = 49$ scenario at higher active user loads. Distributing the antennas makes the system more robust to large-scale fading, resulting in improved performance.

\begin{figure}
    \centering
    \includegraphics[width=1\linewidth]{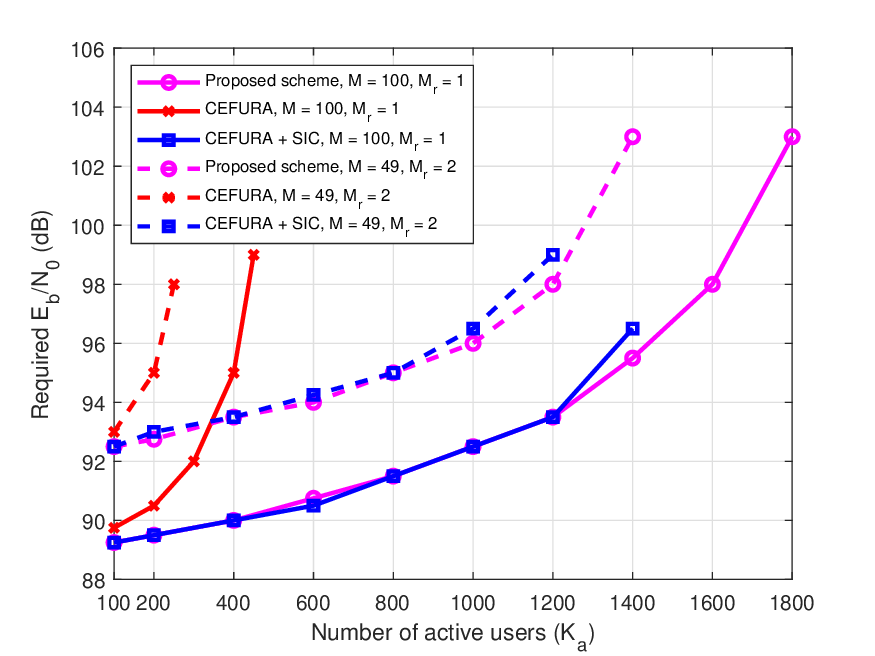}
        \caption{Required transmit $E_b/N_0$ to meet a target PUPE of $P_e \leq 0.05$}
    \label{figeff}
\end{figure}

We evaluate the energy efficiency of the proposed scheme by calculating the required $E_b/N_0$ to achieve a target PUPE of 0.05 and comparing it with that of CEFURA in Fig. \ref{figeff}, over a wide range of active user loads. The required $E_b/N_0$ of the system can be calculated as 

\begin{equation}
    \frac{E_b}{N_0} = \frac{n_pP_p + n_dP_d}{B\sigma^2}.
\end{equation}

Note that the received signal-to-noise ratio is much lower than the transmit energy per bit due to severe path loss and shadowing. The results in Fig. \ref{figeff} demonstrate that the proposed scheme outperforms CEFURA by up to 4.5 dB for $M = 100$ and $M_r = 1$ and 1.5 dB for $M = 49$ and $M_r = 2$ while it significantly increases the number of supported active users, i.e., the maximum number of active users that the system can accommodate while satisfying the target PUPE. Furthermore, even if the CEFURA scheme is modified to add an SIC block, the proposed scheme still offers similar performance with lower complexity and can support more active users. Namely, the proposed scheme can support up to 1800 users for $M = 100$ and $M_r = 1$ and up to 1400 users for $M = 49$ and $M_r = 2$, while CEFURA can support 450 and 250 users for $M = 100$ and $M_r = 1$ and $M = 49$ and $M_r = 2$, respectively, and 1400 and 1200 users for $M = 100$ and $M_r = 1$ and $M = 49$ and $M_r = 2$, respectively, when it is modified by adding an SIC block.

\begin{figure} 
    \centering
    \includegraphics[width=1\linewidth]{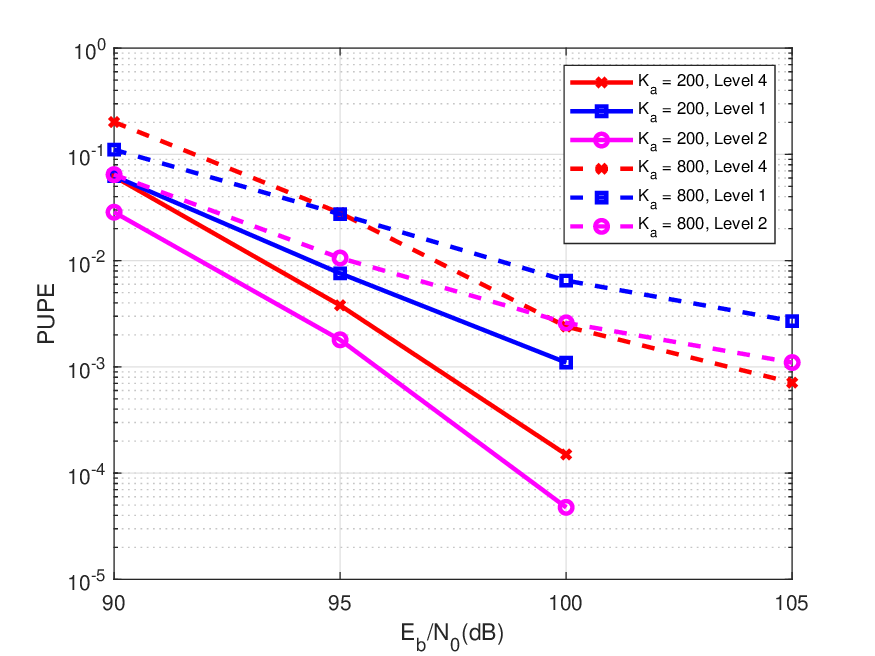}
    \caption{PUPE comparison of different cooperation levels.}
    \label{figcoop}
\end{figure}

\begin{figure} 
    \centering
    \includegraphics[width=1\linewidth]{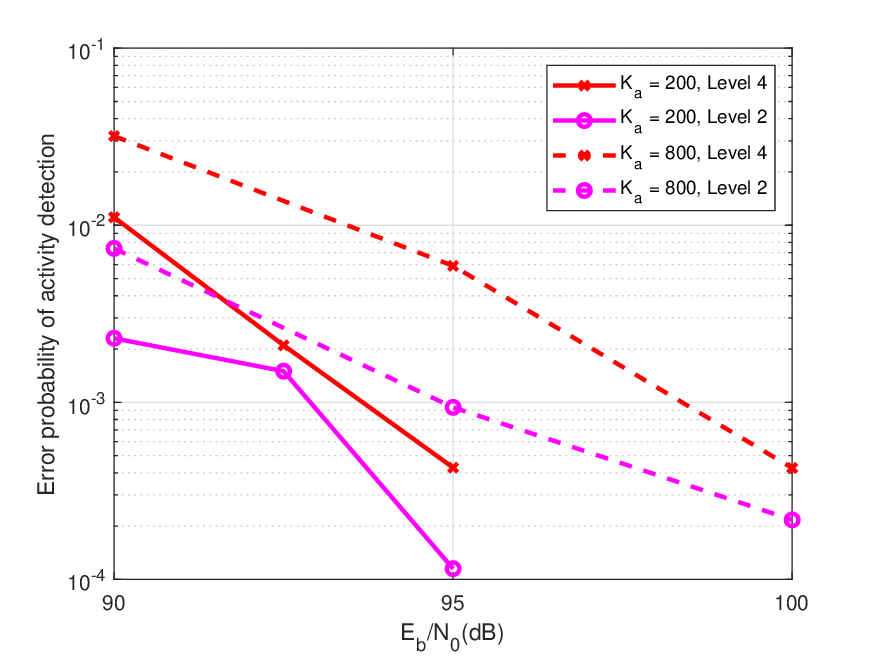}
    \caption{Misdetection probability of OMP for different cooperation levels.}
    \label{figsign}
\end{figure}

We then examine the effect of AP cooperation on the system performance by comparing the PUPE of the \textit{Level 2} implementation, centralized operation (\textit{Level 4}), and the case with no cooperation where each AP tries to recover the message of $K_m$ users without passing the symbol estimates to the CPU (\textit{Level 1}) in Fig. \ref{figcoop}. For this comparison, we take $K_a = 200$ and $800$ and set $M = 100$ and $M_r = 1$. The results in Fig. \ref{figcoop} show that the proposed \textit{Level 2} cooperation outperforms \textit{Level 1} cooperation for both user loads and is superior to the centralized operation (\textit{Level 4} cooperation) for $K_a = 200$. In addition, it outperforms centralized operation for $K_a = 800$ in the high PUPE regime and offers a similar performance otherwise. Although there is more cooperation in the \textit{Level 4} case compared to \textit{Level 2}, the reason that \textit{Level 2} outperforms \textit{Level 4} is as follows: In \textit{Level 4} operation, all users are detected together at the CPU, degrading the activity detection and system performance, as it is lower bounded by the activity detection error. On the other hand, in \textit{Level 2}, a subset of the users are locally detected at each AP, leading to a better activity detection performance as illustrated in Fig. \ref{figsign}. This behavior shows the advantage of \textit{Level 2} operation in massive random access scenarios.

\begin{figure}
    \centering
    \includegraphics[width=1\linewidth]{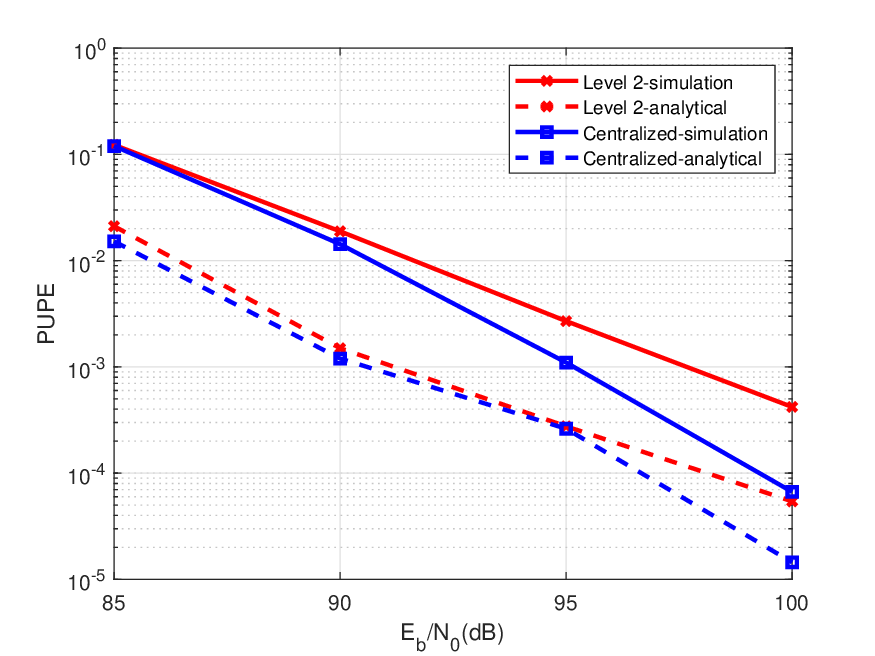}
       \caption{PUPE comparison of analytical and simulation performance of the proposed scheme for $K_a = 50$.}
    \label{figanalysis}
\end{figure}

We verify the normal approximation-based performance analysis in Section \ref{perf} in Fig. \ref{figanalysis}. For this purpose, we perform Monte Carlo simulations to obtain the PUPE value in (\ref{eqerrorfinal}), where $M = 100$, $M_r = 1$, and $K_a = 50$. We set $K_m = 7$ for the \textit{Level 2} scenario. For the actual simulations, we do not account for SIC, since we compute the SINR in the first iteration of the analytical derivations to ensure a fair comparison. The results in Fig. \ref{figanalysis} show that the analytical performance characterization of the proposed scheme captures its actual performance. Namely, the performance gap between the simulation and analytical result is around 5 dB for a PUPE of $10^{-2}$.

\begin{figure}
    \centering
    \includegraphics[width=1\linewidth]{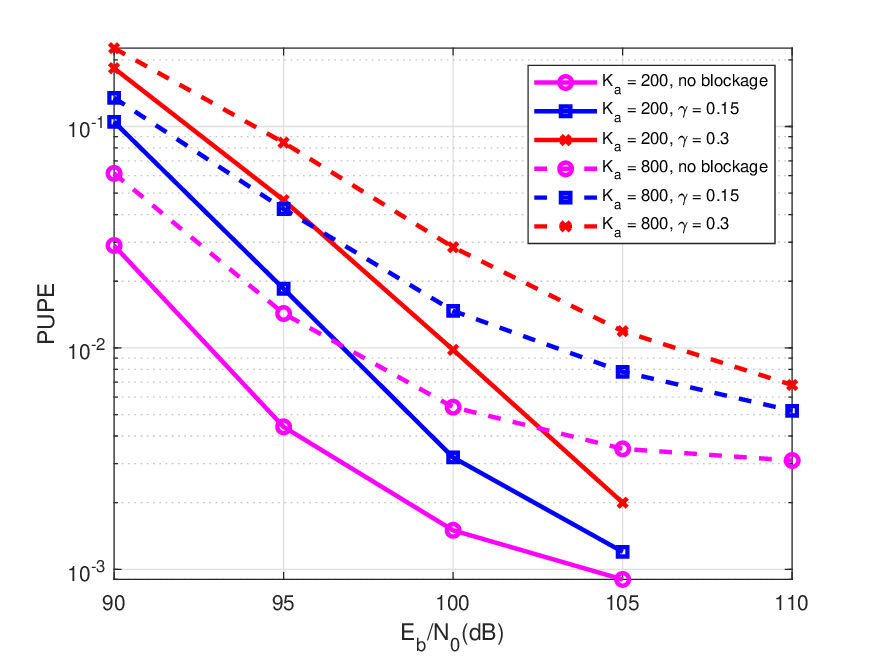}
    \caption{PUPE comparison of proposed scheme with and without blockage for $\gamma = 0.15, 0.3$.}
    \label{figblock}
\end{figure}

We investigate the case where a user's transmission to an AP is blocked with probability $\gamma$, i.i.d. across user-AP pairs in Fig. \ref{figblock}, for $K_a = 200, 800$, and $\gamma = 0.15, 0.3$. In this case, even if its transmission to some APs is blocked, a user can still be detected by other APs, and its message can still be decoded. The results in Fig. \ref{figblock} illustrate that blockage degrades system performance as expected; however, its effect is less pronounced as $E_b/N_0$ increases, especially for $K_a = 200$.

\begin{figure}
    \centering
    \includegraphics[width=1\linewidth]{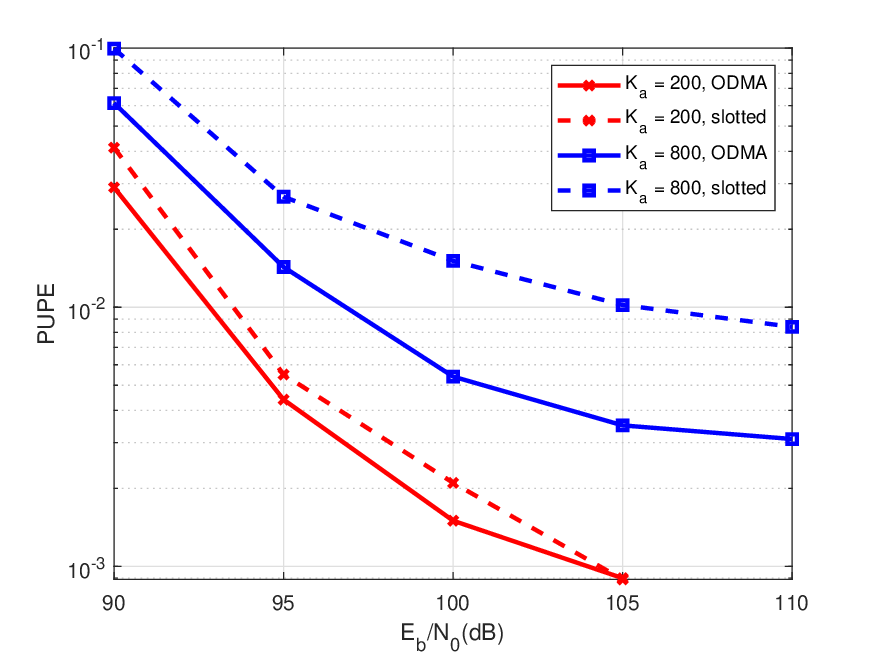}
    \caption{PUPE comparison of ODMA and slotted transmission.}
    \label{figslot}
\end{figure}

Finally, we compare the performance of the ODMA method with that of a state-of-the-art method in URA over MIMO fading channels, namely slotted transmission in \cite{twc}, for $M = 100$ and $M_r = 1$, with $K_a = 200$ and $800$. In slotted transmission, the transmission frame is divided into slots, and each user picks a slot to transmit its signal. At the receiver side, the decoding is performed on a slot-by-slot basis. The results in Fig. \ref{figslot} show that the performance of ODMA and slotted transmission is similar for $K_a = 200$; however, ODMA offers a much better performance for $K_a = 800$. For example, it outperforms slotted transmission by around 8 dB for a PUPE of $10^{-2}$. This demonstrates ODMA's potential under high active-user loads, owing to its structured sparsity and flexibility.

\section{Conclusions} \label{conclusion}

We study URA in a cell-free scenario in which users transmit their signals to distributed APs connected to a CPU via a fronthaul. We propose a scalable and energy-efficient scheme in which the transmission frame is divided into two parts: each active user transmits a pilot sequence in the pilot part, followed by its polar codeword, distributed to the data part according to an on-off pattern. On the receiver side, we employ joint pilot and channel estimation via OMP, LMMSE symbol estimation, symbol combining, and single-user decoding, followed by SIC. We also analyze the system performance using normal approximations. Numerical results illustrate that the proposed scheme outperforms the state-of-the-art scheme in the literature, and the performance predicted by analysis closely matches the simulation results. In this work, we assume that the APs are synchronized; however, an extension of the proposed cell-free URA scheme for the asynchronous setup would be an interesting future direction.

\section*{Appendix}

\subsection*{Calculation of $P_S$, $P_I$ and $P_N$ in Level 2 scenario}

\begin{flalign}
    P_S &= \frac{1}{n_k^2} \mathbb{E} \left[ \sum\limits_{m' =1}^{n_k} \left(\mathbf{{P}}_k^T \mathbf{{P}}_k \mathbf{{c}}_k\mathbf{g}_{k,m'} \mathbf{W}_{m'}^{-1} \mathbf{g}_{k,m'}^H \right)^H \right. \notag \\
    &\quad \left. \sum\limits_{m'' =1}^{n_k} \mathbf{{P}}_k^T \mathbf{{P}}_k \mathbf{{c}}_k\mathbf{g}_{k,m''} \mathbf{W}_{m''}^{-1} \mathbf{g}_{k,m''}^H  \right] \notag \\
    &= \frac{1}{n_k^2} \mathbb{E} \left[  \sum\limits_{m' =1}^{n_k} \sum\limits_{m'' =1}^{n_k} \mathbf{g}_{k,m'} (\mathbf{W}_{m'}^{-1})^H \mathbf{g}_{k,m'}^H  \right.  \\
    &\left. \mathbf{{c}}_k^H \mathbf{{P}}_k^T \mathbf{{P}}_k \mathbf{{P}}_k^T \mathbf{{P}}_k  \mathbf{{c}}_k \mathbf{g}_{k,m''} \mathbf{W}_{m''}^{-1} \mathbf{g}_{k,m''}^H  \right] \notag \\
    &= \scalebox{0.90}{$\frac{1}{n_k^2} n_dP_d \sum\limits_{m' =1}^{n_k} \sum\limits_{m'' =1}^{n_k}  \mathbf{g}_{k,m'} (\mathbf{W}_{m'}^{-1})^H \mathbf{g}_{k,m'}^H \mathbf{g}_{k,m''} \mathbf{W}_{m''}^{-1} \mathbf{g}_{k,m''}^H$}, \notag \\ 
    P_I &= \frac{1}{n_k^2}\mathbb{E} \left[ \sum\limits_{m' =1}^{n_k} \sum\limits_{i =1, i \neq k}^{K_a} \left(\mathbf{{P}}_k^T \mathbf{{P}}_i \mathbf{{c}}_i\mathbf{g}_{i,m'} \mathbf{W}_{m'}^{-1} \mathbf{g}_{k,m'}^H\right)^H \right. \notag \\
    & \left.  \sum\limits_{m''=1}^{n_k} \sum\limits_{j =1, j \neq k}^{K_a} \mathbf{{P}}_k^T \mathbf{{P}}_j \mathbf{{c}}_j\mathbf{g}_{j,m''} \mathbf{W}_{m''}^{-1} \mathbf{g}_{k,m''}^H \right] \notag\\
    &=  \frac{1}{n_k^2}\mathbb{E} \left[ \sum\limits_{m' =1}^{n_k} \sum\limits_{m''=1}^{n_k} \sum\limits_{i =1, i \neq k}^{K_a} \sum\limits_{j =1, j \neq k}^{K_a} \mathbf{g}_{k,m'} \mathbf{W}_{m'}^{-1} \mathbf{g}_{i,m'}^H \mathbf{{c}}_i^H \right. \notag \\
    &\left. \mathbf{{P}}_i^T \mathbf{{P}}_k  \mathbf{{P}}_k^T \mathbf{{P}}_j \mathbf{{c}}_j\mathbf{g}_{j,m''} \mathbf{W}_{m''}^{-1} \mathbf{g}_{k,m''}^H \right] \notag  \\
    &=  \frac{1}{n_k^2} \sum\limits_{m' =1}^{n_k} \sum\limits_{m''=1}^{n_k} \sum\limits_{i =1, i \neq k}^{K_a} \sum\limits_{j =1, j \neq k}^{K_a} \mathbf{g}_{k,m'} \mathbf{W}_{m'}^{-1} \mathbf{g}_{i,m'}^H \notag \\
    &\quad \mathbb{E} \left[\mathbf{{c}}_i^H \mathbf{{P}}_i^T \mathbf{{P}}_k  \mathbf{{P}}_k^T \mathbf{{P}}_j \mathbf{{c}}_j \right] \mathbf{g}_{j,m''}\mathbf{W}_{m''}^{-1} \mathbf{g}_{k,m''}^H \notag \\
    &= \frac{1}{n_k^2} P_d \sum\limits_{i =1, i \neq k}^{K_a}  tr\{\mathbf{{P}}_i^T \mathbf{{P}}_k  \mathbf{{P}}_k^T \mathbf{{P}}_i \} \\
    & \sum\limits_{m' =1}^{n_k} \sum\limits_{m''=1}^{n_k}  \mathbf{g}_{k,m'} \mathbf{W}_{m'}^{-1}  \mathbf{g}_{i,m'}^H \mathbf{g}_{i,m''} \mathbf{W}_{m''}^{-1} \mathbf{g}_{k,m''}^H \notag \\
    &= \scalebox{0.90}{$\frac{1}{n_k^2} P_d \sum\limits_{i =1, i \neq k}^{K_a}  tr\{\mathbf{{P}}_i^T \mathbf{{P}}_k  \mathbf{{P}}_k^T \mathbf{{P}}_i \} \norm{\sum\limits_{m''=1}^{n_k}\mathbf{g}_{i,m''} \mathbf{W}_{m''}^{-1} \mathbf{g}_{k,m''}^H}^2 $}, \notag \\
     P_N &= \scalebox{0.8}{$ \frac{1}{n_k^2} \mathbb{E} \left[ \sum\limits_{m' =1}^{n_k} \left( \mathbf{{P}}_k^T \mathbf{Z}_{m'} \mathbf{W}_{m'}^{-1} \mathbf{g}_{k,m'}^H  \right)^H \sum\limits_{m'' =1}^{n_k} \mathbf{{P}}_k^T \mathbf{Z}_{m''} \mathbf{W}_{m''}^{-1} \mathbf{g}_{k,m''}^H  \right] \notag$} \\
    &= \scalebox{0.9}{$\frac{1}{n_k^2} \mathbb{E} \left[ \sum\limits_{m' =1}^{n_k}  \sum\limits_{m'' =1}^{n_k} \mathbf{g}_{k,m'} \mathbf{W}_{m'}^{-1} \mathbf{Z}_{m'}^H \mathbf{{P}}_k \mathbf{{P}}_k^T \mathbf{Z}_{m''} \mathbf{W}_{m''}^{-1} \mathbf{g}_{k,m''}^H   \right]$} \\
    &= \scalebox{0.9}{$\frac{1}{n_k^2} \sum\limits_{m' =1}^{n_k}  \sum\limits_{m'' =1}^{n_k} \mathbf{g}_{k,m'} \mathbf{W}_{m'}^{-1} \mathbb{E} \left[  \mathbf{Z}_{m'}^H \mathbf{{P}}_k \mathbf{{P}}_k^T \mathbf{Z}_{m''}  \right] \mathbf{W}_{m''}^{-1} \mathbf{g}_{k,m''}^H$} \notag \\
    &= \frac{1}{n_k^2} \sum\limits_{m' =1}^{n_k}   \mathbf{g}_{k,m'} \mathbf{W}_{m'}^{-1}  \mathbb{E} \left[ \mathbf{Z}_{m'}^H \mathbf{{P}}_k \mathbf{{P}}_k^T \mathbf{Z}_{m'} \right] \mathbf{W}_{m'}^{-1} \mathbf{g}_{k,m'}^H \notag \\
    &= \frac{1}{n_k^2} n_d \sigma_z^2 \sum\limits_{m' =1}^{n_k} \norm{\mathbf{W}_{m'}^{-1} \mathbf{g}_{k,m'}^H}^2. \notag 
\end{flalign}

\noindent where the expectation is taken over the transmitted symbols and noise. Note that there is a little bit of abuse of notation, as picking the APs that a user is detected by indicator function is reduced to index $m'$, and $\sum\limits_{m =1}^{M} \mathbbm{1}_{\{ k\in \mathcal{\hat{I}}_m\}}$ is equal to $n_k$.

\subsection*{Calculation of $P_S$, $P_I$ and $P_N$ in Level 4 scenario}

\begin{flalign}
      P_{S,c} &= \mathbb{E} \left[ \left(\mathbf{P}_k^T \mathbf{P}_k \mathbf{c}_k \mathbf{g}_k  \mathbf{W}^{-1} \mathbf{{g}_k}^H \right)^H \mathbf{P}_k^T \mathbf{P}_k \mathbf{c}_k \mathbf{g}_k  \mathbf{W}^{-1} \mathbf{{g}_k}^H \right] \notag \\&= \mathbb{E} \left[  \mathbf{g}_k(\mathbf{W}^{-1})^H \mathbf{{g}_k}^H \mathbf{c}_k^H \mathbf{c}_k \mathbf{g}_k  \mathbf{W}^{-1} \mathbf{{g}_k}^H\right] \\&= n_dP_d \norm{\mathbf{g}_k  \mathbf{W}^{-1} \mathbf{{g}_k}^H}^2, \notag \\P_{I,c} &=  \scalebox{0.75}{$\mathbb{E} \left [ \left(\sum\limits_{j=1, j \neq k }^{K_a} \mathbf{P}_k^T \mathbf{P}_j \mathbf{c}_j \mathbf{g}_j \mathbf{W}^{-1} \mathbf{{g}_k}^H \right)^H \sum\limits_{i=1, i \neq k }^{K_a} \mathbf{P}_k^T \mathbf{P}_i \mathbf{c}_i \mathbf{g}_i \mathbf{W}^{-1} \mathbf{{g}_k}^H  \right]$} \\ &= \scalebox{0.83}{$\mathbf{g}_k \mathbf{W}^{-1} \mathbb{E} \left[ \left( \sum\limits_{j=1, j \neq k }^{K_a} \sum\limits_{i=1, i \neq k }^{K_a} \mathbf{{g}_j}^H \mathbf{c}_j^H  \mathbf{P}_j^T \mathbf{P}_k  \mathbf{P}_k^T \mathbf{P}_i \mathbf{c}_i \mathbf{g}_i\right) \right] \mathbf{W}^{-1} \mathbf{g}_k^H \notag$} \\&= \scalebox{0.9}{$ \mathbf{g}_k \mathbf{W}^{-1} \left( \sum\limits_{j=1, j \neq k }^{K_a} \mathbf{{g}_j}^H tr\{\mathbf{P}_j^T \mathbf{P}_k  \mathbf{P}_k^T \mathbf{P}_j\} n_dP_d \mathbf{{g}_j}  \right) \mathbf{W}^{-1} \mathbf{g}_k^H$} \notag \\ &= \scalebox{0.95}{$P_d \mathbf{g}_k \mathbf{W}^{-1} \left(\sum\limits_{j=1, j \neq k  }^{K_a} tr\{\mathbf{P}_j^T \mathbf{P}_k  \mathbf{P}_k^T \mathbf{P}_j\} \mathbf{{g}_j}^H \mathbf{{g}_j}   \right) \mathbf{W}^{-1} \mathbf{g}_k^H $}, \notag \\P_{N,c} &=  \mathbb{E} \left[ \left(\mathbf{P}_k^T \mathbf{Z}_d \mathbf{W}^{-1} \mathbf{{g}_k}^H \right)^H \mathbf{P}_k^T \mathbf{Z}_d \mathbf{W}^{-1} \mathbf{{g}_k}^H \right] \notag \\ &= \mathbb{E} \left[ \mathbf{{g}_k} \mathbf{W}^{-1} \mathbf{Z}_d^H \mathbf{P}_k \mathbf{P}_k^T \mathbf{Z}_d \mathbf{W}^{-1} \mathbf{{g}_k}^H\right]  \\ &= \mathbf{{g}_k} \mathbf{W}^{-1} \mathbb{E} \left[ \mathbf{Z}_d^H \mathbf{P}_k \mathbf{P}_k^T \mathbf{Z}_d  \right] \mathbf{W}^{-1} \mathbf{{g}_k}^H \notag \\ &= n_d \sigma_z^2 \norm{\mathbf{W}^{-1} \mathbf{{g}_k}^H}^2. \notag
\end{flalign}

\end{document}